\documentclass[12pt,a4]{article}

\usepackage[dvips]{graphics}
\usepackage{epsfig}

\newcommand{\e}{\varepsilon}
\newcommand{\Fix}{\mbox{Fix}}
\newcommand{\id}{id}

\begin{document}

\title{Dynamical Networks in Function Dynamics}
\author{Naoto Kataoka\footnote{Corresponding author. kataoka@aurora.es.hokudai.ac.jp (N.Kataoka).} and Kunihiko Kaneko\\
  {\small \sl Meme Media Laboratory (VBL), Faculty of Engineering,}\\
  {\small \sl Hokkaido University, N12 W6 Sapporo, 060-0812, JAPAN}\\
  {\small \sl Department of Pure and Applied Sciences,}\\
  {\small \sl University of Tokyo, Komaba, Meguro-ku, Tokyo 153, JAPAN}\\
}

\date{}
\maketitle

\begin{abstract}
{\small As a first step toward realizing a dynamical system that evolves
while spontaneously
determining its own rule for time evolution,
function dynamics (FD) is analyzed.
FD consists of a functional equation with a self-referential term,
given as a dynamical system of a 1-dimensional map.
Through the time evolution of this system,
a dynamical graph (a network) emerges.  
This graph has three interesting properties:
i)   vertices appear as stable elements,
ii)  the terminals of directed edges change in time, and 
iii) some vertices determine the dynamics of edges, 
     and edges determine the stability of the vertices, complementarily.
Two aspects of FD are studied, 
the generation of a graph (network) structure 
and the dynamics of this graph (network) in the system.\\

}
\end{abstract}

\section{Introduction}

In studying biological, linguistic, and social systems,
one is generally interested in understanding how a rule for the temporal evolution of 
a system is organized through the dynamics of the system itself.
To construct a model of such a system as a dynamical system or
to simulate it using a computer, however,
we need to supply a rule for the temporal evolution of the system.
It would thus seem that the goal of understanding the spontaneously emergence of 
such a rule in biological systems through such modeling is difficult to achieve. 

For example, in a dynamical systems approach to the modeling of such a system, 
a rule governing the evolution of  a set of variables takes the form of a set of equations.
Then, in a naive attempt to realize the goal stated above,
we may introduce a dynamical system describing the evolution of
these equations, as a `rule forming' dynamics.
However, the resulting `dynamical dynamical system' would be nothing but
a high-dimensional dynamical system.
With the same goal in mind, then,
we would need to introduce another dynamical 
system to describe the change undergone by this high-dimensional system.
We thus see how a naive attempt to realize the spontaneous
emergence of a fundamental dynamics-governing rule results in an infinite hierarchy of systems
and necessarily meets with failure.

In this paper, we consider a model
with the potentiality to form such a hierarchy spontaneously 
within an infinite-dimensional dynamical system.
We will show that low-dimensional dynamical systems
are generated from the original infinite-dimensional dynamical
system. 
The generation of such low-dimensional systems constitutes the generation
of a dynamic-governing rule.
We elucidate the mechanism for this rule generation.

To serve the present purpose of studying biological systems, 
The function dynamics (FD) that we previously
introduced \cite{NKKI}\cite{NKKII}\cite{TKKN} is reformulated
in Sec.2.  In this FD, we adopt a
functional equation having a `self-reference' term,
i.e., a term including composition of a function (say $f\circ f$).
With this term, it is shown that a rule governing the dynamics of partial 
functions restricted to certain intervals is generated.
This rule is determined  
by partial functions on
other intervals.  With this relationship among partial functions,
low-dimensional dynamical systems are formed from the
original infinite dimensional dynamical system.
Accordingly, the function dynamics of the original system 
can be interpreted as dynamics over partial functions. 
From this point of view, the partial functions can be considered elements in a self-organized,
low-dimensional system that evolves together while mutually generating rules that govern each other's
dynamics.
We first demonstrate that function dynamics generates rules 
hierarchically among partial functions.
Then, we find that the relationships 
among partial functions are sometimes entangled, in the sense that
one partial function gives a rule for some other 
partial functions, which give a rule for the original partial function.
The formation of successive rules for low-dimensional systems and 
this entanglement of rules is analyzed.

In Sec.3, formation of low-dimensional dynamical systems is studied, and
in the Sec.4, the dynamics of the low-dimensional system are investigated.
A summary and discussion are given in Sec.5.

\section{The Model and Its Basic Features}

Let $f$ be a 1-dimensional map, $I \to I$.
We define a map from the 1-dimensional map $f$ to a new 1-dimensional map
as $\Psi_{\e}(f) = (1-\e) id + \e f$.
Here $\id$ is an identity function and $\e$ is a constant ($0 < \e < 1$).
This map is well defined if $f(I)\subseteq I$ is satisfied.
We define a FD system employing such $f$ and $\Psi$ as follows:
\begin{equation}\label{eq:1}
f_{n+1} = \Psi_{\e}(f_n)\circ f_n  = (1-\e)f_n + \e f_n\circ f_n.
\end{equation}
This equation determines the evolution of the function $f_n$.
If the initial function $f_0$ satisfies $f_0(I) \subseteq I$,
the temporal evolution, i.e. $f_n$ ($n=1,2,\cdots$), is determined.

The evolution equation of a function is generally an infinite-dimensional
dynamical system.  Among the class of systems consisting of such function dynamics, 
the existence of the `self-referential' term $f_n\circ f_n$ is 
the distinguishing feature that characterizes the system (1).
This composition term $f_n\circ f_n$ determines the temporal evolution of $f_n$,
and it is an operation in the functional space. With this term,
the rule governing the evolution of a functional value at a particular value of its argument
depends on the values of the function taken at other values of the argument.

\begin{figure}[htbp]
\begin{center}
\begin{tabular}{ll}
(a) &(b)\\
\includegraphics[width=6cm, height=6cm]{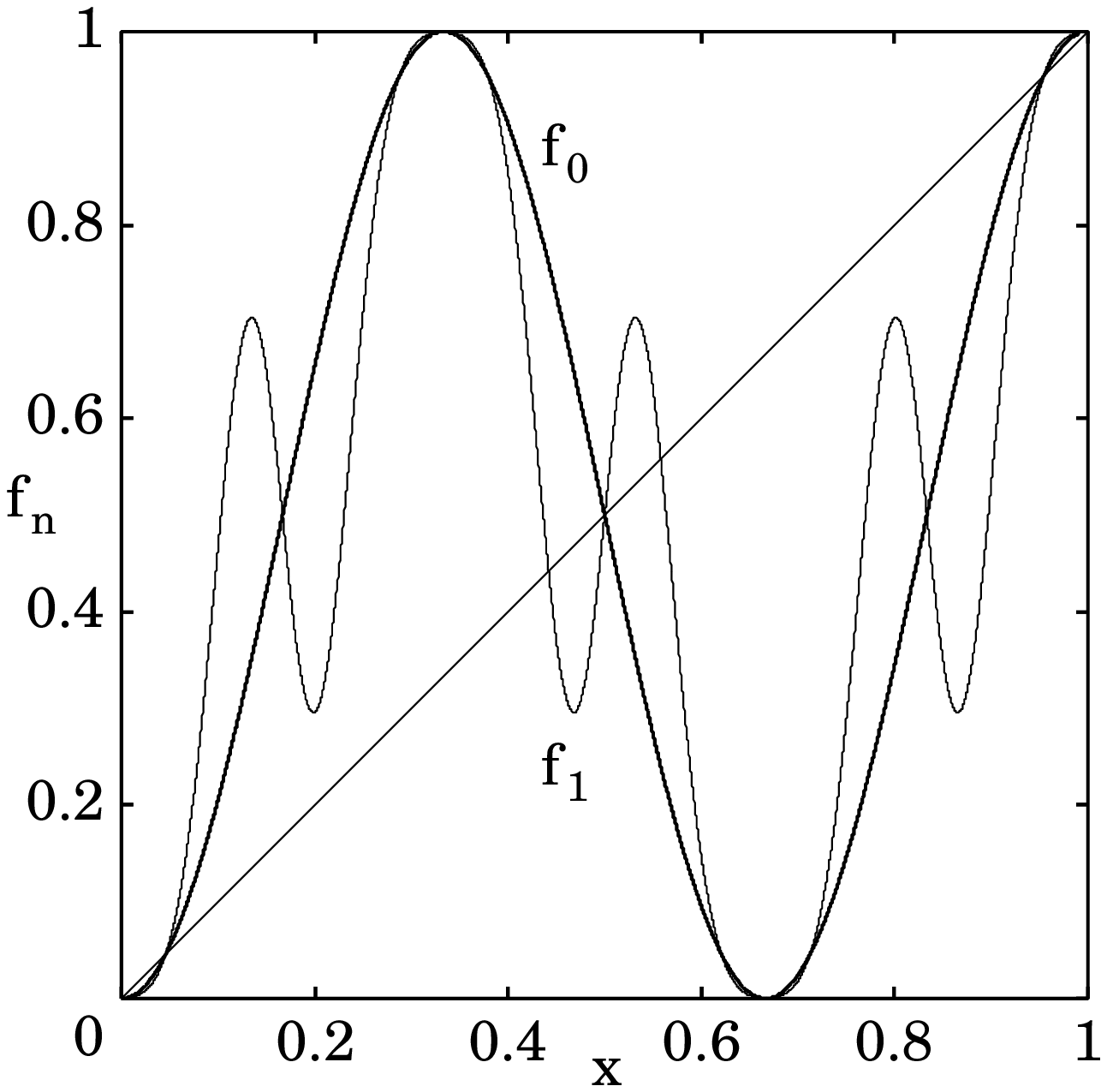} &
\includegraphics[width=6cm, height=6cm]{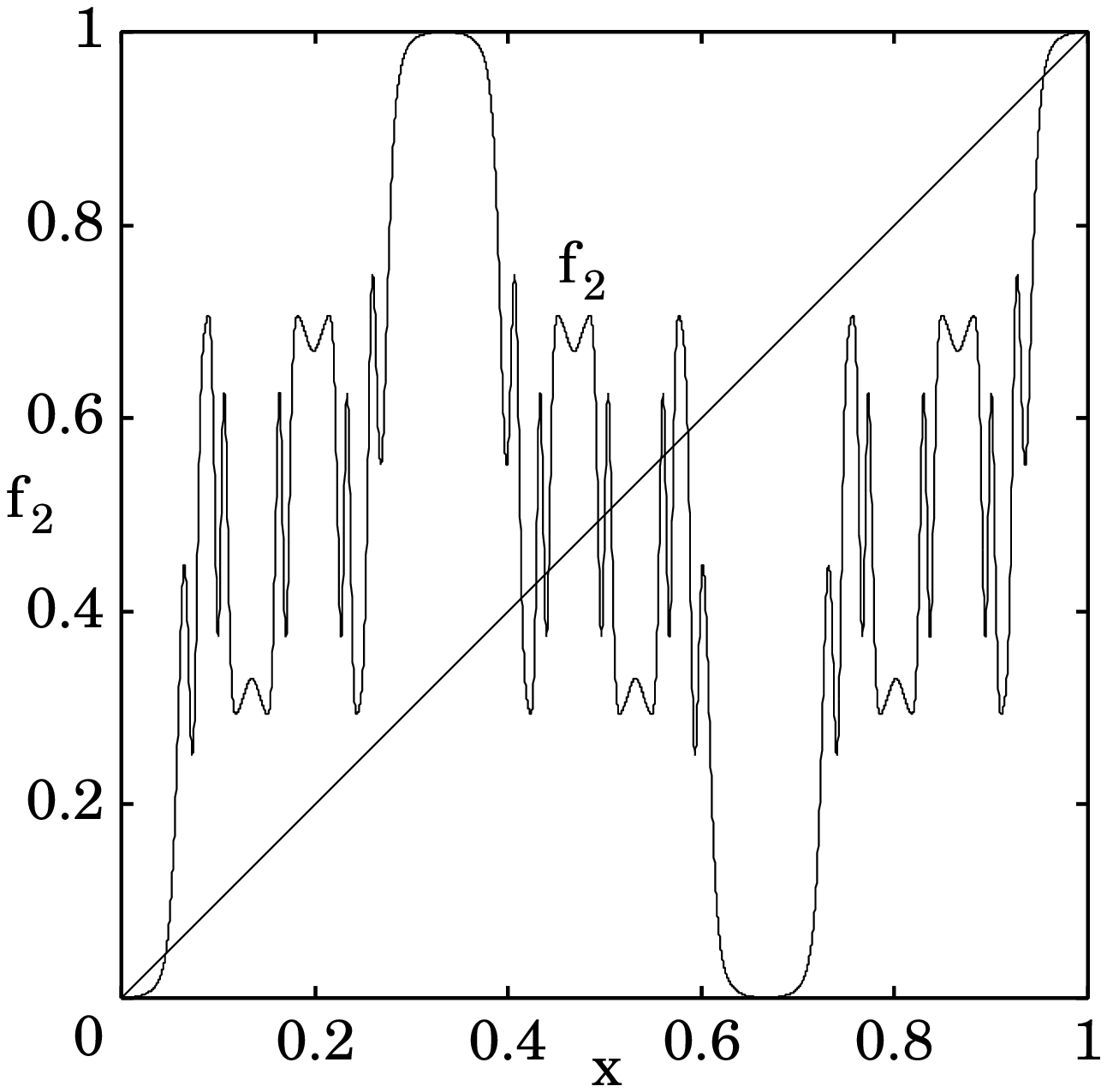}\\
(c) &(d)\\
\includegraphics[width=6cm, height=6cm]{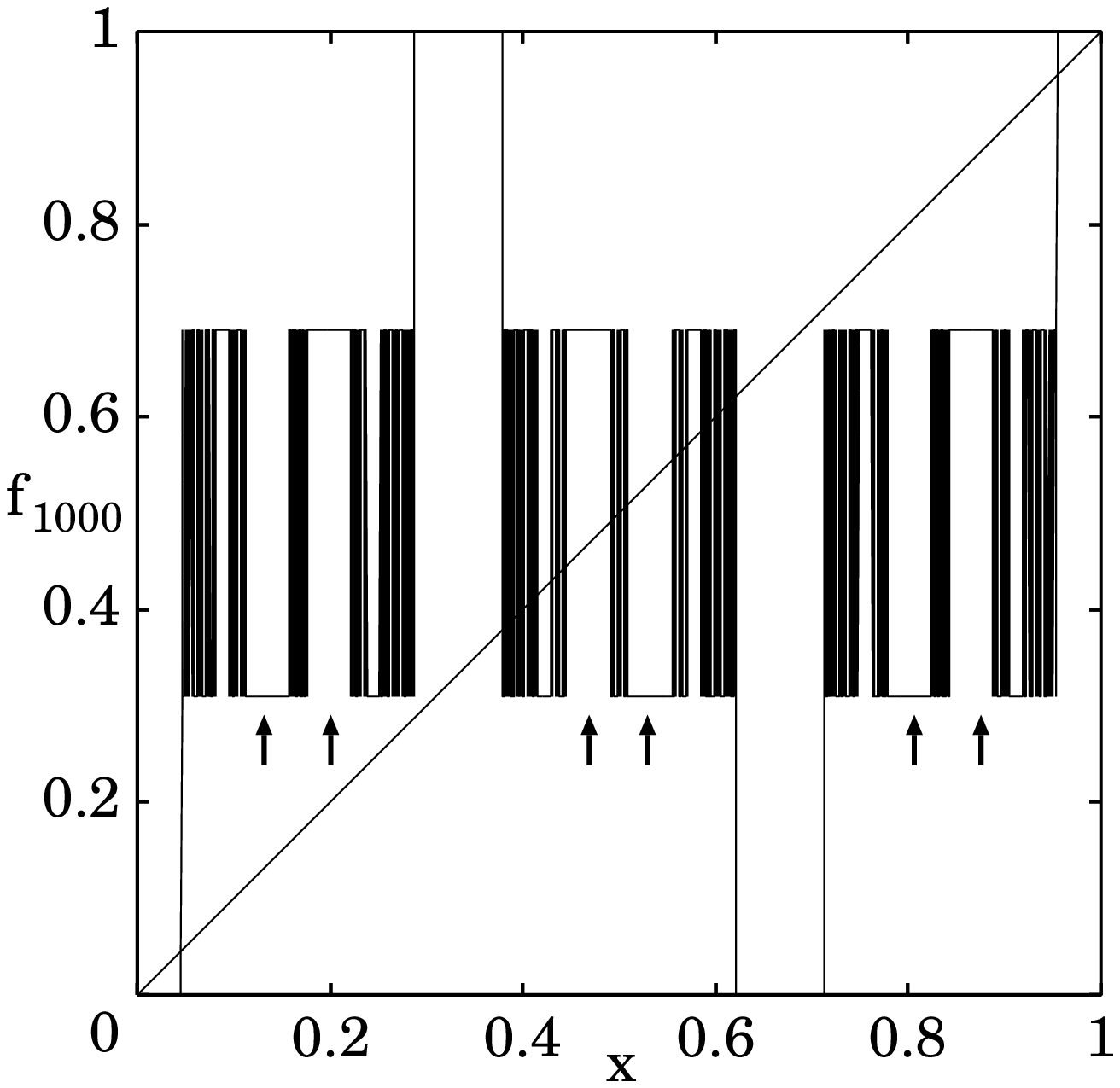}&
\includegraphics[width=6cm, height=6cm]{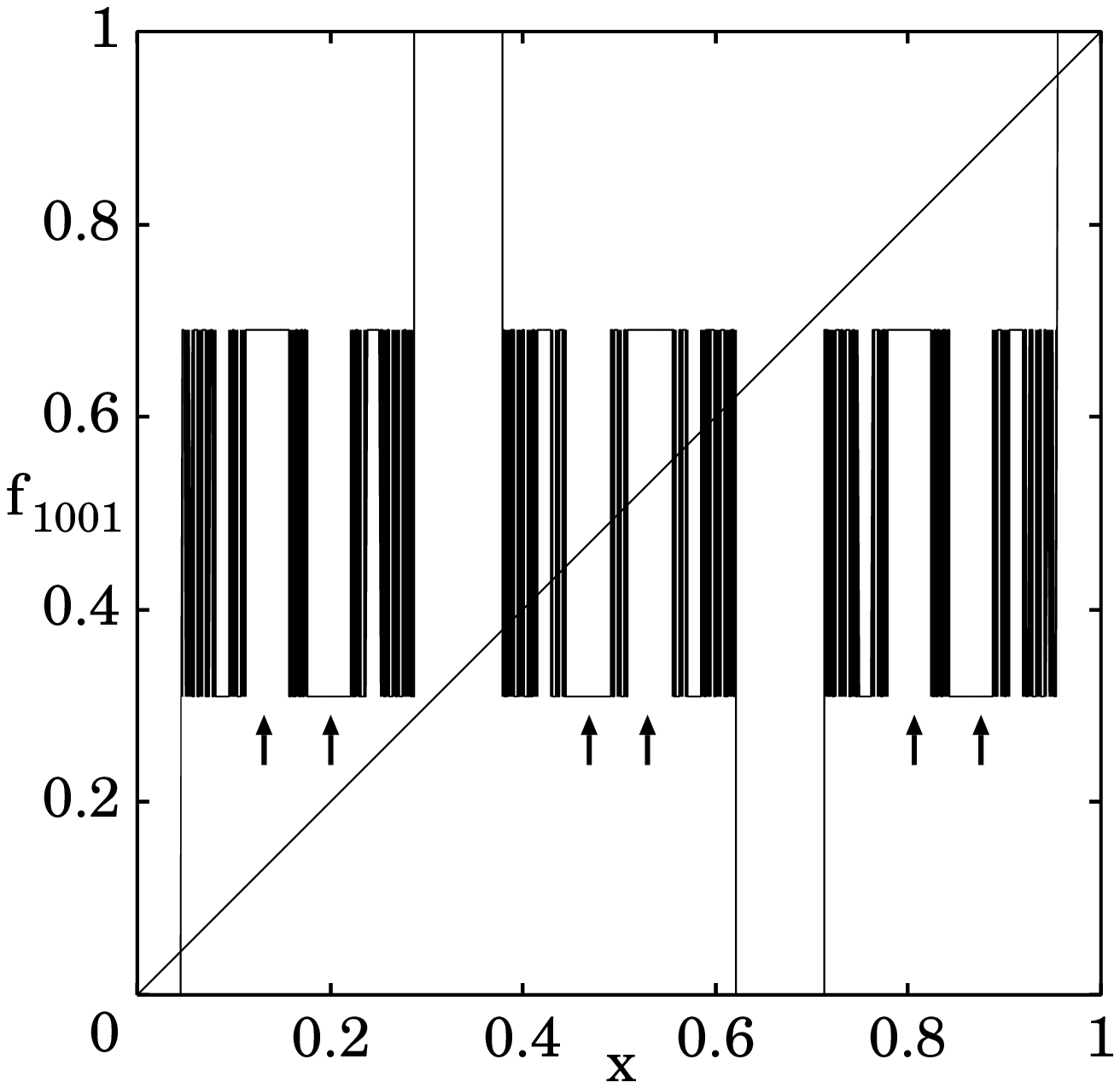}\\
\end{tabular}
\caption{\label{fig:sin_evolution} {\small An example of the time evolution described by (1)
with $f_0(x) = \sin^2(1.5\pi x)$ and $\e = 0.55$.
Graphed here are (a) $f_0$ and $f_1$, 
(b) $f_2$,
(c) $f_{1000}$ and
(d) $f_{1001}$.
The mesh size is $M = 4000$.
In this example $f_{\infty}$ is on a period-2 attractor in the functional space.
The arrows in (c) and (d) are included to make the period-2 oscillation easily recognizable.  
Later, the graph of $f_n$ oscillates between the forms shown in (c) (for even $n$)
and (d) (for odd $n$).
}
}
\end{center}
\end{figure}

Next, we give an example of the temporal evolution of $f_n$ determined by (1).
Here we set the initial function $f_0(x) = \sin^2(1.5\pi x)$ and $\e = 0.55$.
In Fig.\ref{fig:sin_evolution}, the graph of $f_n$ for certain values of $n$ are displayed.
We find that after a transient, $f_n$ exhibits a temporal oscillation with a period of 2 time steps.
In this example, the total function falls on a period-2 attractor in 
the functional space.
However, for some other values of $\e$, more complicated behavior appears.
In general, it is not easy to fully explore all the phenomena
displayed by our model by computer simulation alone.

To carry out a numerical computation for the above FD, it is necessary to divide an interval into 
a finite number, $M$, of mesh points.  For example, $I = [0, 1)$ could be divided into subintervals as
$I^i = [i/M, (i+1)/M)$ ($i = 0, \ldots, M-1$).
However, the composition term often causes a folding of the graph of the 
function at each time step, as shown in Figs.1 (a) and (b).
For an initial function with a single hump, 
the number of foldings can increase approximately in proportion to $2^n$.
Hence, finer and finer structures can be generated
in the graph of $f_n$, and as a result the outcome of the simulation can have a strong dependence on 
the mesh size $M$.
Because the folding effect generates infinitesimal (Cantor-set-like)
structures \cite{YH} with a variety of scales, 
it is difficult to study this system by direct computer simulation.\\

In general, the following relations are fundamental for the study of FD:
\begin{itemize}
\item If $f_n(x') = f_n(x'')$ is satisfied at some $n$, 
then $f_m(x') = f_m(x'')$ $\forall m>n$
\item A fixed point of  $f_n$ [$f_n(q) = q$] is $n$-independent 
[i.e., $f_n(q) = q$ implies $f_m(q)$ $\forall m>n$].
\end{itemize}
We denote the set of fixed points of a 1-dimensional map $f$ by $\Fix (f)$.
Note that a point $x' \not \in \Fix (f_n)$ satisfying $f_n(x') = q \in \Fix(f_n)$ 
is also $n$-independent in the FD.
In other words, 
if the value of $f_n$ at a point is equal to a fixed point, then this point is also $n$-independent.
The set of all such $n$-independent points [including all points in \Fix(f)] is denoted $\Omega(f)$.

\section{Network Structure} 

In Figs.\ref{fig:sin_evolution}(c) and (d),
each $f_n$ seems to be a piecewise constant function (that is, 
the graph of $f_n$ consists of flat pieces).
This property is not peculiar to this initial function, but is rather general.
Although we have simulated Eq.(1) using a variety of initial functions,
in all cases, the function $f_n$ for sufficiently large $n$
consists of a set of 
constant partial functions. 
(Strictly speaking, $f_n$ approaches such a form asymptotically as $n\rightarrow \infty$.)
Thus we find through our computer simulations
that the number of elements of the set\footnote{Here we denote the number of elements of a set $A$ as $|A|$.}
$\{f_n(i/M)\}$
($i = 0, \ldots M-1$, with $n$ fixed), $|\{f_n(i/(M-1))\}|$, 
satisfies $|\{f_n(i/(M-1))\}|\ll M$ for large $n$,
in the computer simulation \cite{NKKI}.

In this section the generation of flat pieces in the graph of $f_n$ is studied.
Because FD consists of dynamics of a 1-dimensional map, 
to avoid
confusion between the function on the interval $I$ and 
that on the functional space, we define some useful notation.

\begin{itemize}
\item A ``partial function $f_n|_A$'' is a function defined on an interval $A$ ($A \subset I$), 
and $f_n|_A (x) = f_n(x)$ for $x \in A$.
\item A ``fixed point of $f_n$'' means a fixed point of the 1-dimensional map $f_n$. 
A ``fixed point of the FD'' means a 1-dimensional map 
$f_n$ satisfying that $f_{n+1} = f_n$.
\item If a partial function $f_n|_A$ satisfies 
$f_m|_A = f_n|_A$ $\forall m>n$,
we say that ``$f_n|_A$ is $n$-independent''.
\item The statement ``$f_n|_A$ is a constant function'' means that 
$f_n(A) = c$ (where $c$ is a real number) for fixed $n$.
It is important to keep distinction between the statements ``$f_n|_A$ is a constant function''
and ``$f_n|_A$ is $n$-independent'' clear.
\item The term ``temporal evolution'' refers to that of the function $f_n$ (i.e., $f_0, f_1, \ldots$).
It does not refer to the dynamics of the 1-dimensional map, given by
$x_{m+1} = f_n(x_m)$.
\item We define a network for the function $f_n$, in order to describe a graph of $f_n$ in terms of
graph theory. In general, a network consists of elements (vertices) and edges. 
Throughout the paper, we use $A$ in reference to connected entire intervals 
(i.e., intervals of constant $f_n|_A$ that are not proper subset of any other such intervals).
We consider each interval $A$ as an element (i.e., vertex).
We consider a directed edge to be associated with $f_n|_A$ and $A$ and denote it $f_n|_A\rightarrow A$.
A dynamical ``network'' consists of elements formed through time evolution 
and edges whose terminals change in time.
\end{itemize}

\subsection{Stability around a Fixed Point}

Now we discuss how a constant partial function is formed in FD and study
its stability.
First, we consider an initial function 
$f_0(x) = a_0(x - q) + q$, around the fixed point $q$. 
The evolution equation of the slope $a_n$ is easily found to take the form
$a_{n+1} = (1-\e)a_n + \e {a_n}^2$.
This 1-dimensional map has  fixed points $a_n = 0$ and $1$, 
with $a_n=0$ stable and $a_n=1$ unstable.
The basin of the fixed point $0$ is $(-1/\e, 1)$.
Thus if ${f_n}'(x)$ is constant and the relation $-1/\e< {f_n}'(x) < 1$ is satisfied 
on $A$ around a fixed point, then
$a_n = {f_n}'(x) \rightarrow 0$ as $n \rightarrow \infty$ and hence $f_{\infty}(x) = q$.
Thus, the slope of the graph of $f_n|_A(x)$ goes to 0.

The following theorem has been proved \cite{TKKN}.

{\bf Theorem.} Consider the function  $d(x) = \frac{-1}{\e} (x - q) + q$.
If $f_n(x)$ is continuous around a fixed point $q$ (on the interval $q \in A\subseteq I$) 
and satisfies the conditions

\noindent (i)  $(d(x) - f_n(x))(x - f_n(x)) < 0$ for $x \in A\setminus\{q\}$,

\noindent (ii) $f_n(A) \subseteq A$,

\noindent then
$f_n|_A(x) \rightarrow q$ as $n\rightarrow \infty$.
\begin{figure}[htbp]
\begin{center}
  \includegraphics[width=6cm, height=6cm]{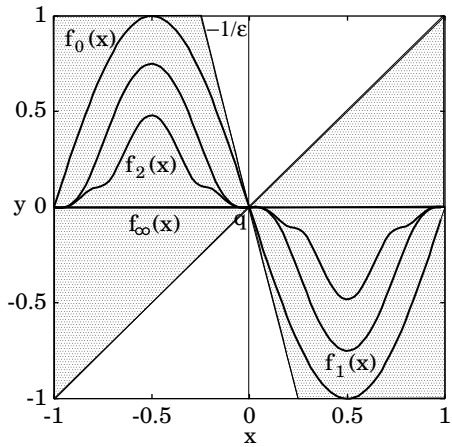}
  \caption{\label{fig:stability} {\small 
If the graph of a continuous function is contained within the shaded area 
around the fixed point $q$ in the figure,
the function converges to a constant $f_{\infty}(x) = q$.
Here, $f_0(x) = -\sin(\pi x)$ $(q = 0)$ and $\e = 1/4$.
}
}
\end{center}
\end{figure}

In Fig.\ref{fig:stability}, the assertion of this theorem is depicted graphically.

As $n$ increases, the number of fixed points of $f_n$ increases 
(see Fig.\ref{fig:sin_evolution}).
If the conditions stated in the theorem are satisfied
around a fixed point,
the graph of the function around this fixed point becomes flat as $n$ increases.
However, this does not necessarily imply that
$f_n$ always converges to a piecewise constant $n$-independent 
function.  We give counterexamples
in the next subsection.

\subsection{Generated Map}

We investigated the FD using a `generated map'
in earlier papers \cite{NKKII}\cite{TKKN}.
This generated map is defined as 
\begin{equation}\label{eq:2}
g_n= \Psi(f_n) = (1-\e)\id + \e f_n.
\end{equation}
This map is determined by $f_n$.  
(We use the term `generated' here because
$g_n$ is generated from $f_n$.)
Using this map, the FD (1) can be represented by 
$f_{n+1} = g_n(f_n)$.
For each $x$, the value $f_{n+1}(x)$ is determined by the equation
$f_{n+1}(x) = g_n|_A(y)$ (where $y = f_n(x)$ and $y \in A$).
Also note that the value $g_n|_A$ is determined by $f_n|_A$.
(In the following discussion, 
we also use the term `drive', as in the phrase `$f_n(x)$ is driven by $g_n|_A$',
and the term `refer', as in `$f_n(x)$ refers to $A$'.)

\begin{figure}[htbp]
\begin{center}
  \includegraphics[width=11cm, height=6cm]{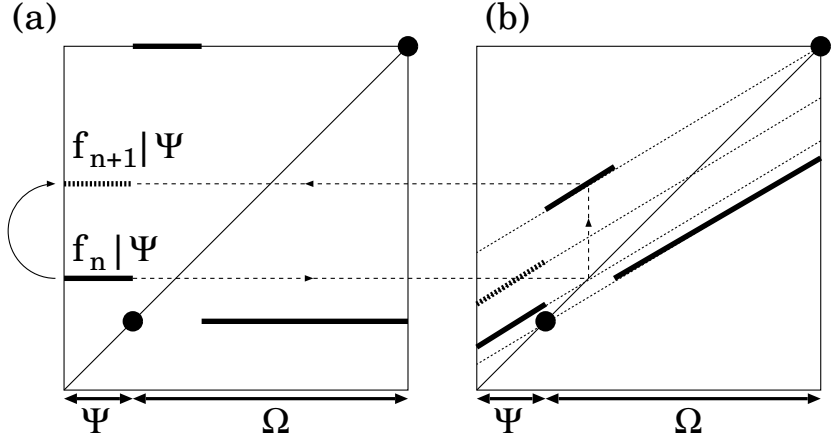}
  \caption{\label{fig:1} {\small Examples of a graph of (a) $f_n(x)$ and 
(b) its generated map $g_n(x)$.
Here $f_n|_{\Omega}(x) \in \Fix(f_n)$ and therefore
$f_n|_{\Omega}$ is an $n$-independent partial function.
$f_n(\Psi) \subset \Omega$, and thus $f_{n+1}|_{\Psi} = g_n|_{\Omega} \circ f_n|_{\Psi}$.
The slope of $g_n|_{\Omega}$ is $1-\e$. In this example, $\e = 3/5$.
The thin broken line crosses the identity function at $f_n(x)$ and has slope $1-\e$.
}
}
\end{center}
\end{figure}

Let us rewrite the $g_n(x)$ in the form $g_n(x) = (1-\e)(x - f_n(x)) + f_n(x)$.
This form is useful to prepare a suitable initial function $f_0$.
If $f_n(A) = c$, then
$g_n|_{A} (x) = (1-\e)(x - c) + c$.
The graph of this generated map is a segment contained in the line of slope $1-\e$ 
that crosses the identity function at $c$.
Examples of (a) a graph of $f_n(x)$ and (b) the corresponding generated map $g_n(x)$ are displayed in Fig.\ref{fig:1}.
For $f_n(x)\in A$, $f_{n+1}(x)$ is determined by $f_{n+1}(x) = g_n|_A\circ f_n(x)$.
If the attractor\footnote{Here the word `attractor' is used in a rather broad sense, because 
$g_n$ can be an $n$-dependent function.} 
of the generated map $g_n$ is not a fixed point, an
$n$-dependent function $f_n(x)$ can exist.

Choosing a suitable $f_0$, it is possible for the map $g_n$ generated 
from an $n$-independent $f_n|_{\Omega}$ to have a 
periodic attractor.  The typical generated map on $\Omega(f_n)$ is
a piece-wise linear map of slope $1-\e$.  The most general one
is given by the Nagumo-Sato map,
$$
x_{n+1} = (1-\e) x_n + w \bmod 1, 0 < \e < 1.
$$
This map is generated from the initial function,
\begin{equation}\label{eq:3}
f_0(x) =
\left\{
\begin{array}{ccl}
\frac{w}{\e}   & \mbox{for} &x \in [0, \frac{1-w}{1-\e})\cup \{\frac{w}{\e}\}\\
\frac{w-1}{\e} & \mbox{for} &x \in [\frac{1-w}{1-\e}, 1)\cup \{\frac{w-1}{\e}\}.
\end{array}
\right.
\end{equation}
Note that each partial function is constant here.

\subsection{Emergence of Elements}

In the Subsection 3.1, we stated a theorem asserting the stability around a fixed point, and 
in Subsection 3.2, we introduced the generated map.
To close the present section, 
we now study the emergence of a piecewise 
constant function.

If $f_n(x)$ is constant around a fixed point $q$ (on $A = [q-a, q+b]$),
the corresponding generated map is given by $g_n|_A(x) = (1-\e)(x-q) + q$.
This generated map crosses the identity function at $q$ with  slope $1-\e$ and
has a stable fixed point $q$.
Thus, if there is a subinterval $B$ ($B\cup A=\emptyset$) on which 
$f_n(B) \subseteq A$ is satisfied,
then $f_n(B) \rightarrow q$ as $n \rightarrow \infty$, and a new flat part is generated.
This constant function is $n$-independent (i.e. $A, B\subseteq \Omega$).

Since the generated map of a constant partial function $f_n|_{\Phi}$
\footnote{This interval $\Phi$ is not necessarily identical to $\Omega$.}
has slope $1-\e$.
As long as the values $f_n(x')$ and $f_n(x'')$ refer to $\Phi$,
the difference between them decreases by the factor $1-\e$ per iteration.
That is, if $f_n(x'') = f_n(x') + \delta$, and $f_n(x')$ and $f_n(x'')$ refer to $\Phi$,
then $|f_{n+1}(x') - f_{n+1}(x'')| = (1-\e) |f_n(x'') - f_n(x') | = (1-\e) \delta$.
From this  consideration, 
$f_{\infty}$ is expected to be a piecewise constant function.
In fact, all computer simulations carried out to this time support this conjecture.\footnote{The 
partial function $f_0(x)|_A=x$ is also $n$-independent.
If we start from this identity partial function, it is maintained under the 
FD.  However, a slight perturbation from it leads to a piecewise constant
function.}.

From the graph theoretical point of view,
the function $f_n$ is regarded as a graph with vertices $x'$ and
 directed edges $x' \leftarrow f_n(x')$.\footnote{Note that the direction 
of an edge is opposite to the direction of the mapping (see Sec.4.2).}
In FD, a vertex corresponding to a flat piece is regarded as a single `element',
while an edge changes in time,
according to the dynamics of $f_n(x')$.
The evolution rule for edges is determined by the generated map, which is determined by the elements.
In the next section, a class of network dynamics exhibited by FD systems is studied.

\section{Dynamical Network}

In the previous section, the emergence of `elements' was discussed.
There, it was seen how as $n\rightarrow \infty$, 
a function $f_n$ approaches a piecewise constant function, with each constant piece regarded as
an element in a network.  Each constant partial function determines  
one linear part of a piecewise linear generated map with slope $(1-\e)$.
The generated map determines the dynamics of the edges.
In this section, we study these dynamics.

\subsection{Choice of the Initial Function}

As illustrated for a particular case in Fig.\ref{fig:sin_evolution}, 
we have found that beginning from most continuous initial functions,
$f_n$ evolves torward a limit function with a shape that is too complicated to be computed.
As long as we are interested in complex dynamics, however,
we need to study the rather  
complicated generated maps that are formed generally through such temporal evolution.

We start by considering computer simulations to obtain a very rough sketch of the functional space.
\begin{figure}[htbp]
\begin{center}
\begin{tabular}{ll}
(a-1) & (a-2)\\
\includegraphics[width=5cm,height=5cm]{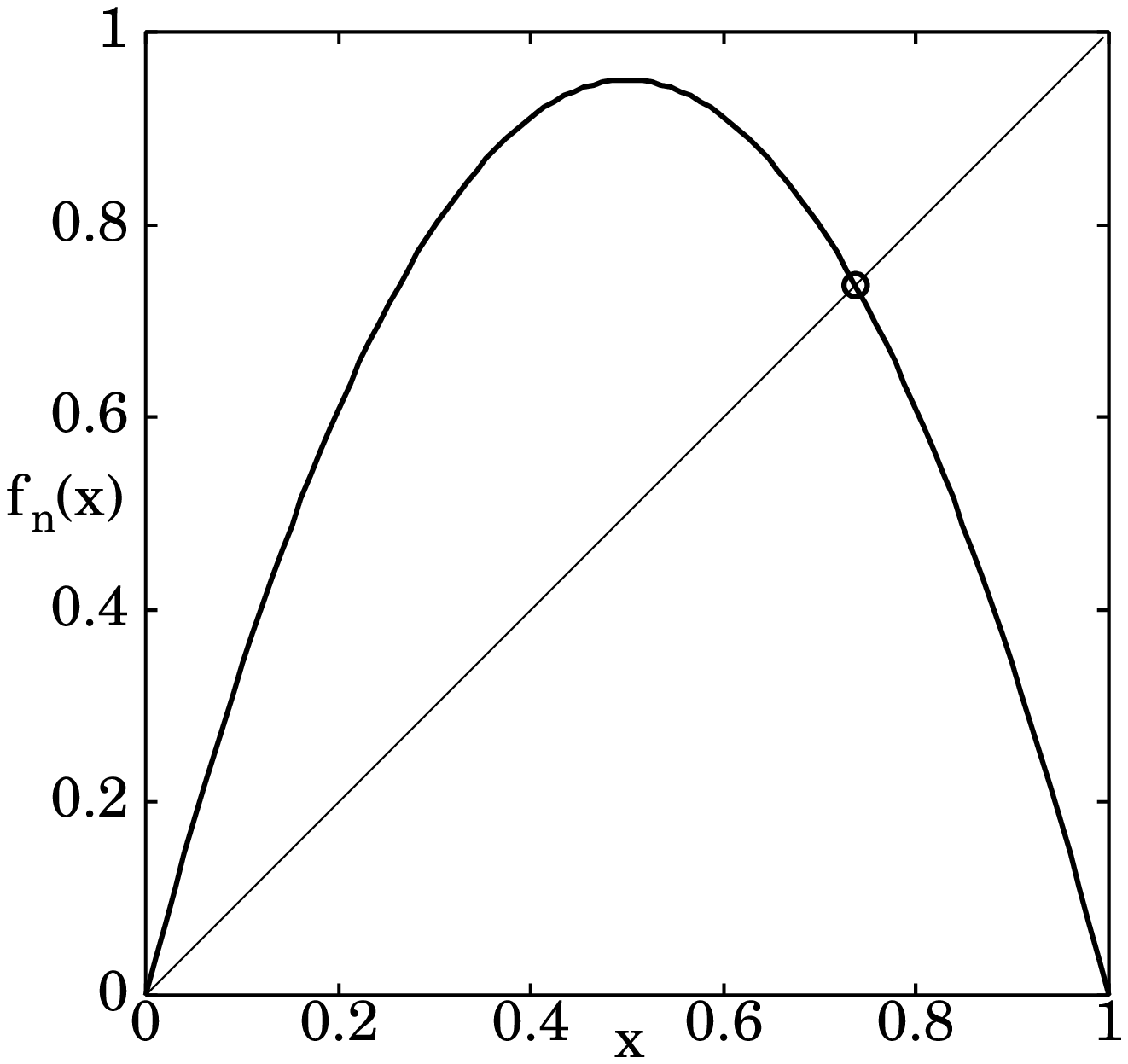}&
\includegraphics[width=5cm,height=5cm]{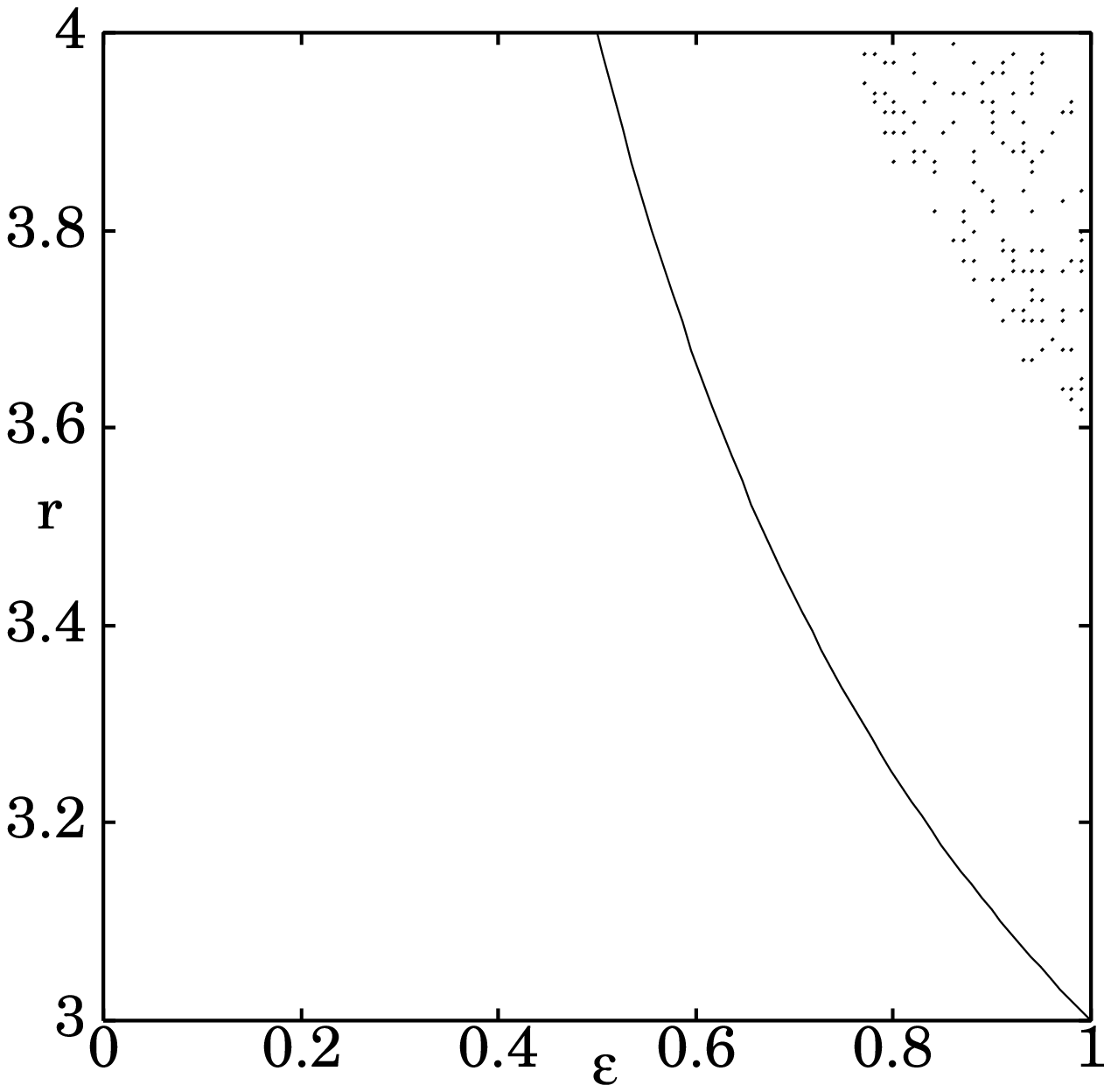}\\

(b-1) & (b-2)\\
\includegraphics[width=5cm,height=5cm]{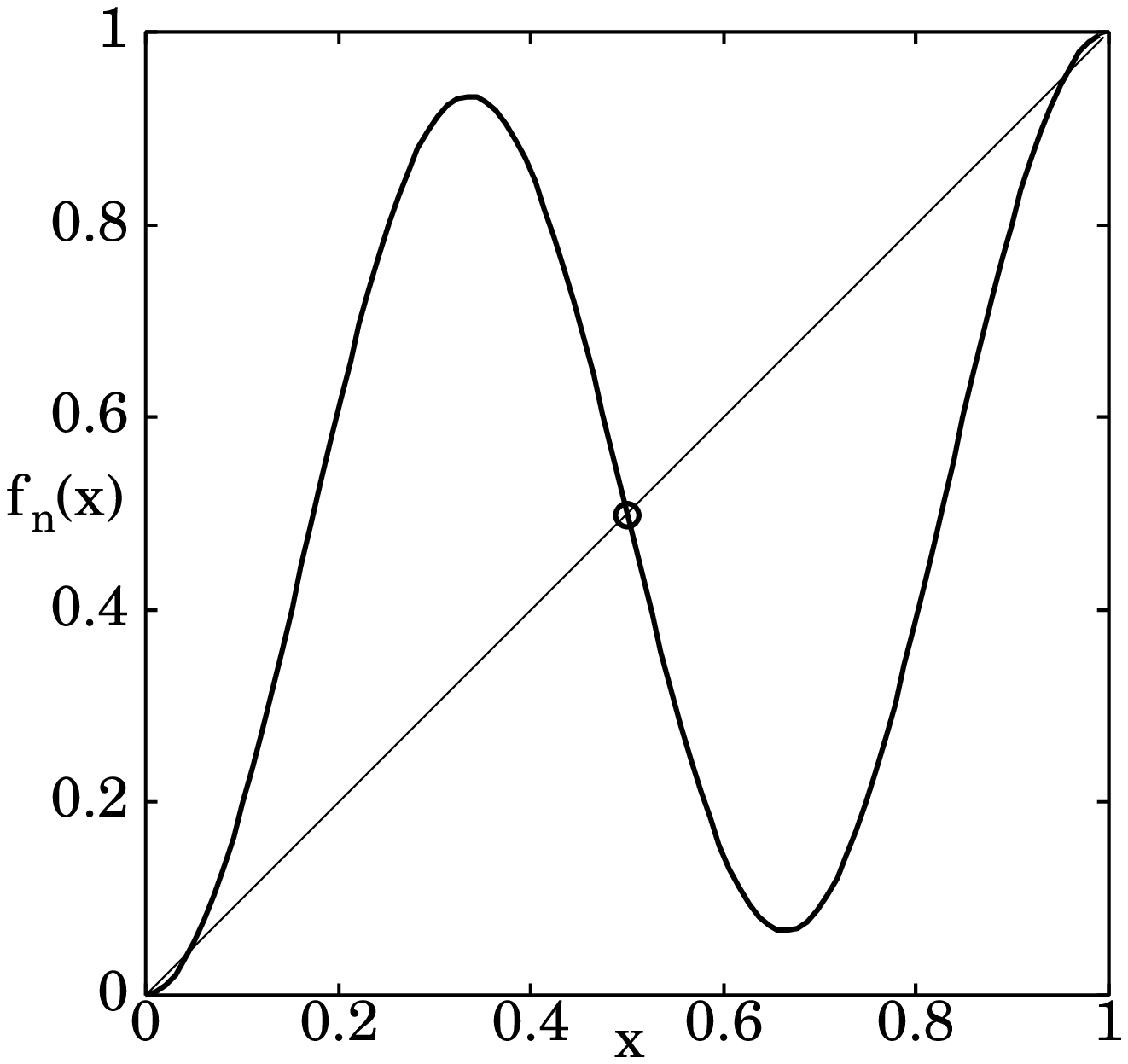} &
\includegraphics[width=5cm,height=5cm]{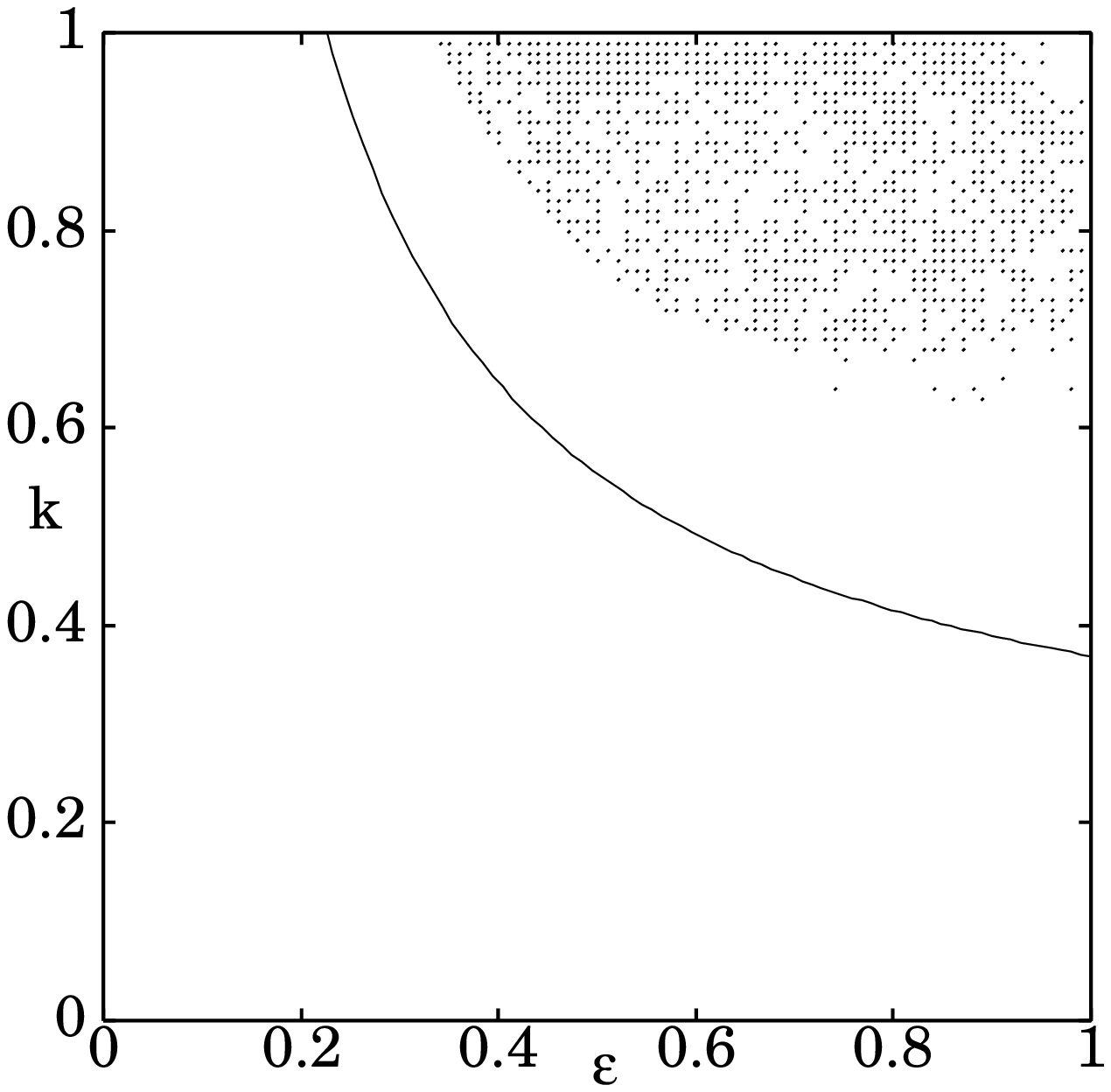}\\

(c-1) & (c-2)\\
\includegraphics[width=5cm,height=5cm]{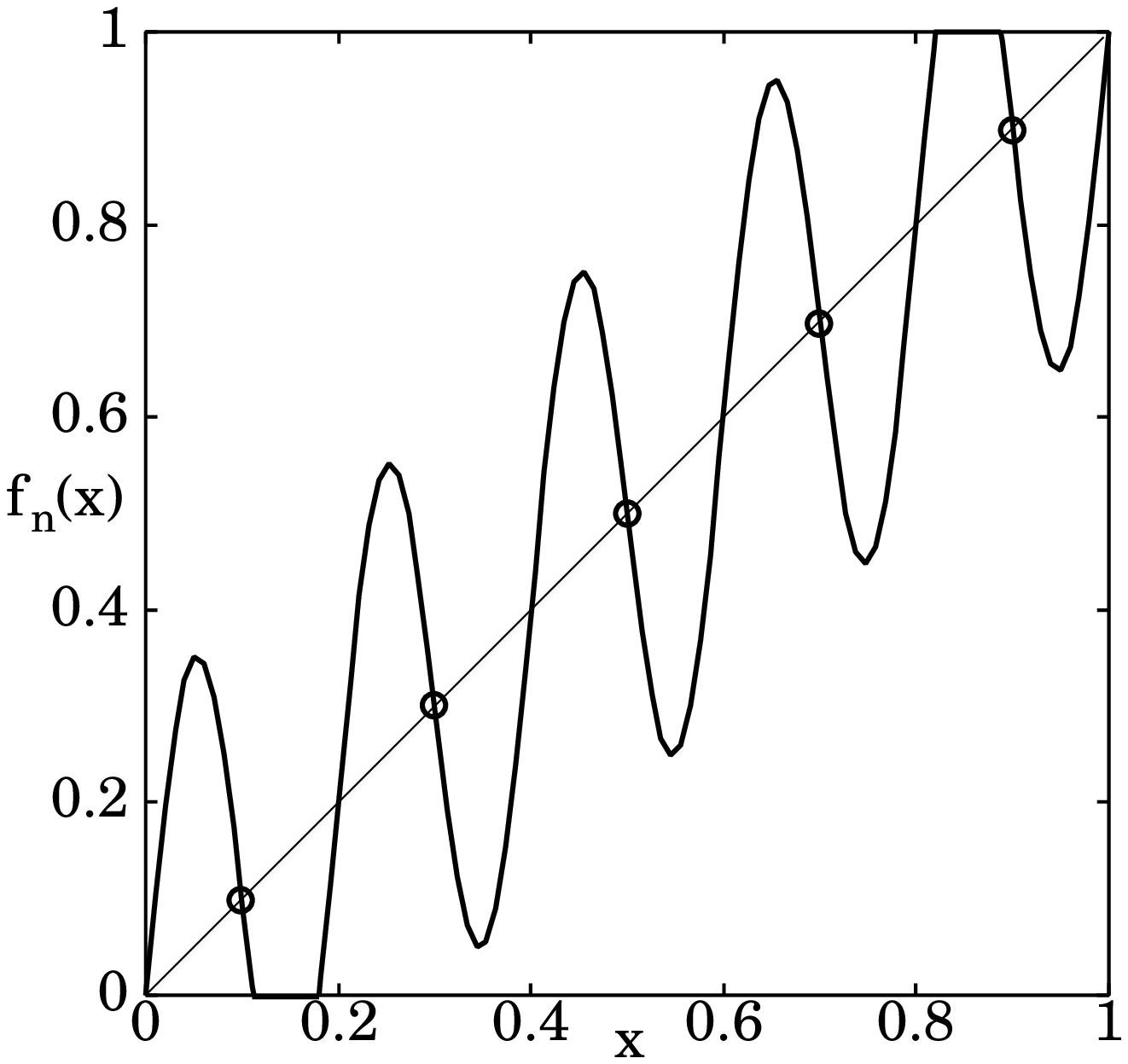} &
\includegraphics[width=5cm,height=5cm]{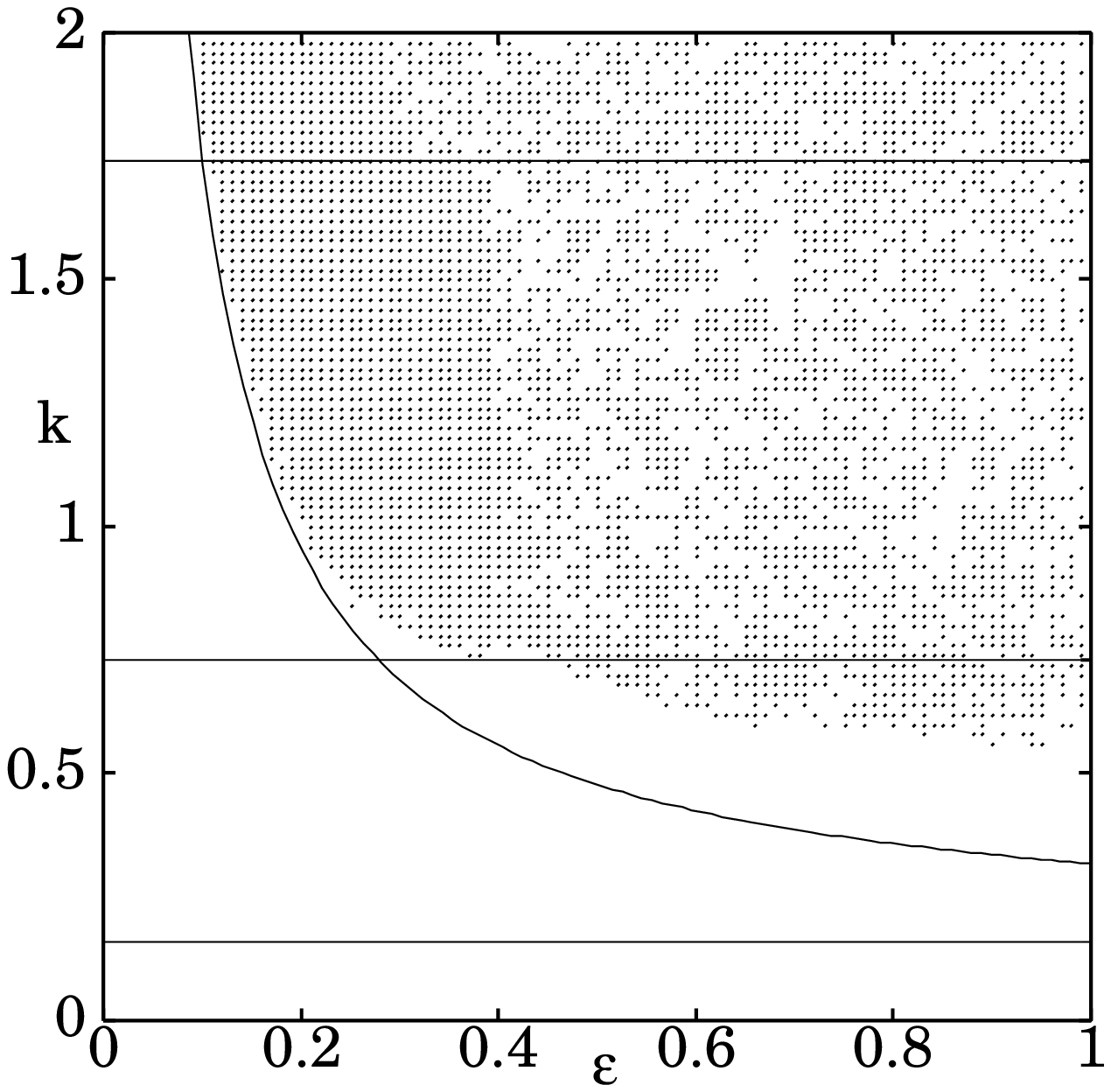}\\
\end{tabular}
\caption{\label{fig:unstable}{\small
Phase diagram ($\cdot$-2) regarding the behavior of the attractor of $f_n$, starting from
the initial functions displayed in ($\cdot$-1), obtained through numerical
simulations using the mesh number
$M = 6000$. Dots are plotted at points
($r,\e$) in (a) and ($k,\e$) in (b) and (c) in the case that $f_n$ does not converge after 1000 steps.
In (a-1), $f_0(x) = rx(1-x)$, $r = 3.8$. 
In (a-2), $u(x) = 2+ \frac{1}{\e}$.
In (b-1), $f_0(x) = (1-k) x + k \sin^2(\frac{3}{2} \pi x)$, $k = 0.9$. 
In (b-2), $u(x) = \frac{2}{\sqrt{2} + 3\pi}(1+\frac{1}{\e})$.
In (c-1), $f_0(x) = x + \frac{k}{m}\sin(2\pi m x)$ $[\mbox{if}$ $f_0(x) < 0 , f_0(x) = 0,$
while $\mbox{if}$ $f_0(x) > 1 , f_0(x) = 1]$, $m = 5$, $k = 1.5$. 
In (c-2), $u(x) = \frac{1}{2\pi}(1+\frac{1}{\e})$. 
}
}
\end{center}
\end{figure}
In Fig.\ref{fig:unstable}, we display three initial functions ($\cdot$-1) 
and the corresponding rough phase diagrams obtained numerically ($\cdot$-2).
In these phase diagrams ($\cdot$-2), the horizontal axis represents $\e$,
and  the vertical axis represents a parameter characterizing $f_0(x)$,
which becomes a steeper as the value of this parameter increases.
In this diagram, a dot is plotted in the case that $f_n$ does not converge to an $n$-independent 
function after 1000 steps (which is generally sufficient for transient behavior to have died out).  
In other words, for a parameter set indicated by a dot,
the attractor of $f_n$ is time dependent, while at all other points,
$f_n$ converges to an $n$-independent function.

Recall that the stability around a fixed point depends on $\e$ (see Sec.3.1).
The stability criterion curve given in Sec.3.1 is also plotted in
Figs.4.($\cdot$-2).  Below this curve,
the fixed points indicated in ($\cdot$-1) are stable (while above the curve they are unstable), 
and therefore here the graph around the fixed point converges to flat pieces.

If the graph around a fixed point $q$ ($q \in A$) is flat, 
then all $f_n(x) \in A$ converge to $q$ as $n \rightarrow \infty$.
As shown in Fig.\ref{fig:unstable}, the stability 
around a fixed point plays an important role in determining whether $f_n$ is 
$n$-dependent or $n$-independent.\footnote{As shown previously, 
there is a gap between the stability criterion curve of the
fixed point and the boundary between the phases
corresponding to $n$-dependent and $n$-independent functions.  Probably this gap can be
reduced by considering the fixed points of $f_1$, 
$f_2$, $\ldots$, successively, and by using a proper
renormalization procedure \cite{Feigenbaum}. However, this has not yet
been successfully carried out.}

To avoid the complicated effect of the folding of the graph of $f_n$,
and to focus on the dynamics of the network, 
we choose an appropriate piecewise constant function as the initial function $f_0$.
Let $I$ be divided into $N$ non-overlapping subintervals $I^i$
($I = \cup_0^{N-1} I^i$ and $I^i \cap I^j= \emptyset$).
The initial function is chosen as 
\begin{equation}\label{eq:4}
f_0(x) = \sum_{i = 1}^N {a^i}_0 1^{I^i} (x).
\end{equation}
Here, the indicating function is defined as
$$
1^{I^i}(x)=
\left\{
\begin{array}{ll}
1 & x \in I^i\\
0 & x \not\in I^i,
\end{array}
\right.
$$
and $f_0(I^i) = {a^i}_0$.
In this section, we consider the case in which the ${a^i}_0$ are chosen randomly,
using a uniform distribution on $[0,1]$.

Although the above initial function may seem too simple or too special
to study the network dynamics of a FD system,
in fact this is not the case.
\begin{itemize}
\item
As mentioned in the previous section, 
a continuous $f_n$ approaches a piecewise constant function 
(consisting of a finite or infinite number of pieces)
as $n \rightarrow \infty$.
\item
To study the dynamics of $f_n$, 
determining the stability around a fixed point is important.
If a continuous partial function is unstable on both sides of a fixed point $q$ for all $n$,
then the value of the partial function does not converge to $q$.
In such a situation, we believe that partial functions $f_n|_{[a, q)}$ and $f_n|_{(q, b]}$
approaches a constant function and do not take the value $q$.
\end{itemize}

Of course, the set of limit functions resulting from the set of all continuous initial functions 
is not identical to that resulting from the set of all piecewise constant initial functions.
However, the limit function realized with a piecewise constant initial function takes 
finite number of values (if $N$ is finite),
and each partial function displays periodic motion with finite period (because ${g_n}' = 1-\e< 1$).
This implies that there exists a continuous initial function $h_0$ that
satisfies  $h_0|_A = f_n|_A$ 
on the set $A = \{f_n(I^i)\}$ (for all $i$ after the transient behavior has died out).

\subsection{Classification of the Dynamical Network}

\begin{figure}[htbp]
\begin{center}
\begin{tabular}{ll}
(a-1) & (a-2)\\
\includegraphics[width=6cm,height=6cm]{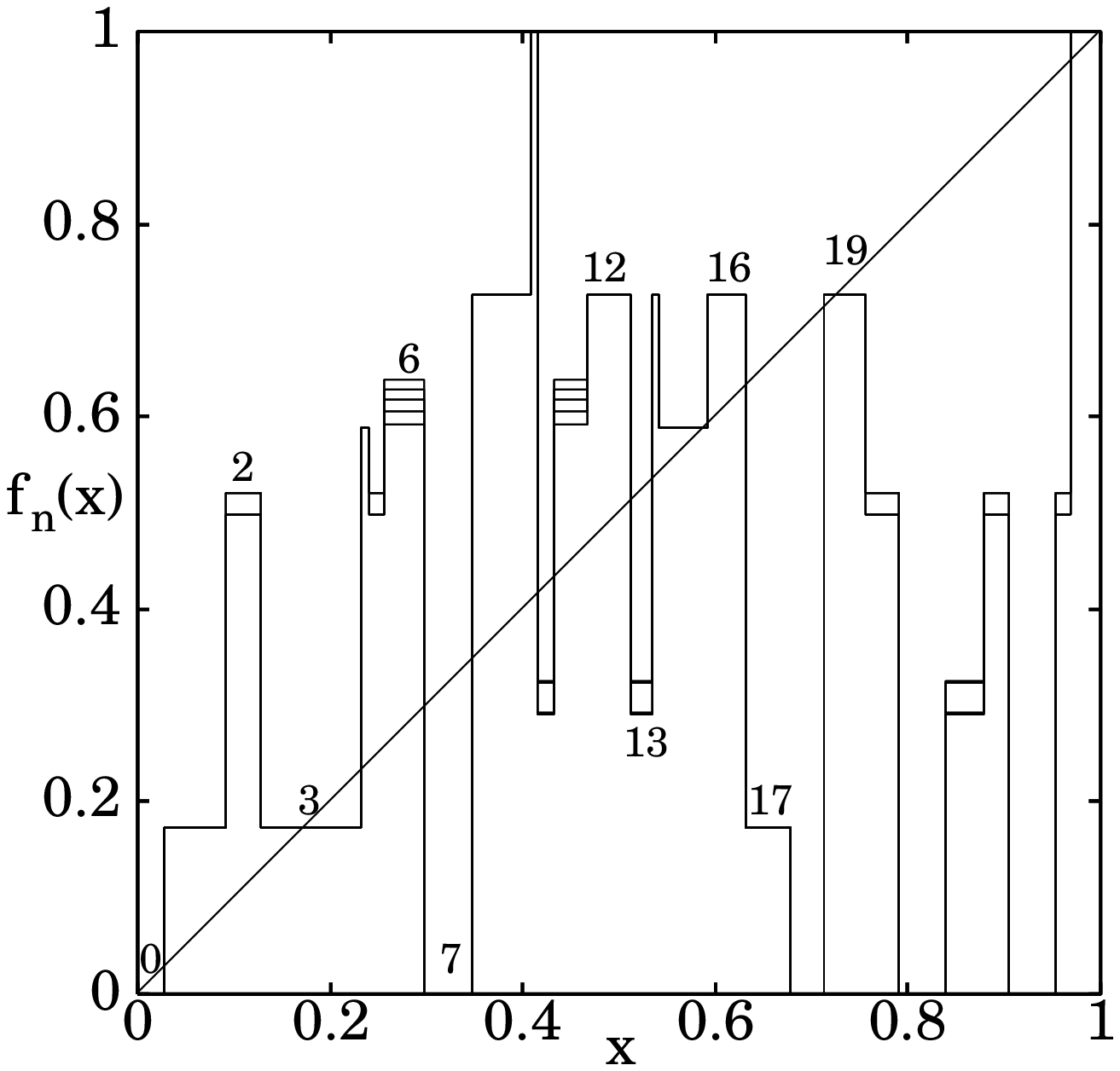}&
\includegraphics[width=6cm,height=6cm]{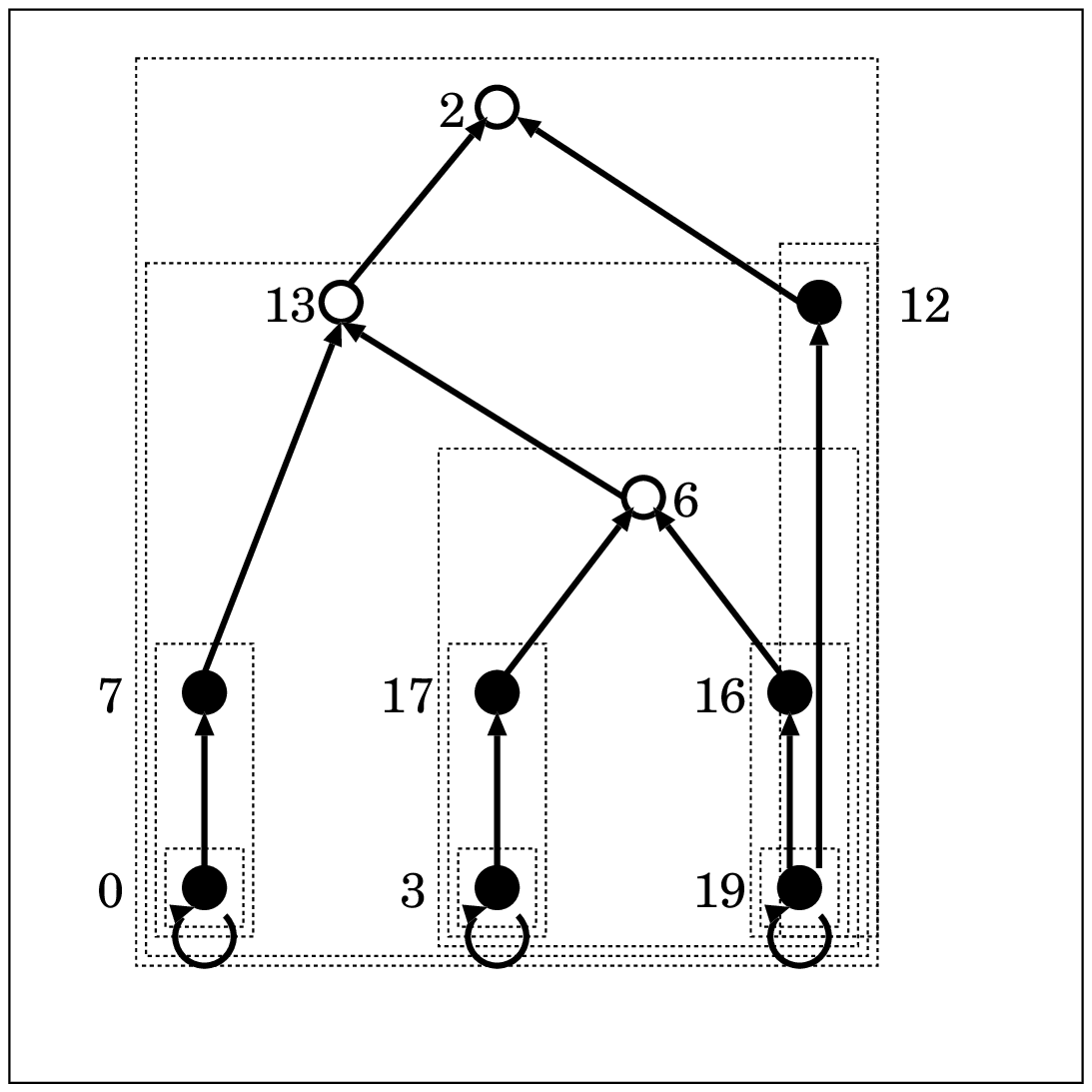}\\

(b-1) & (b-2)\\
\includegraphics[width=6cm,height=6cm]{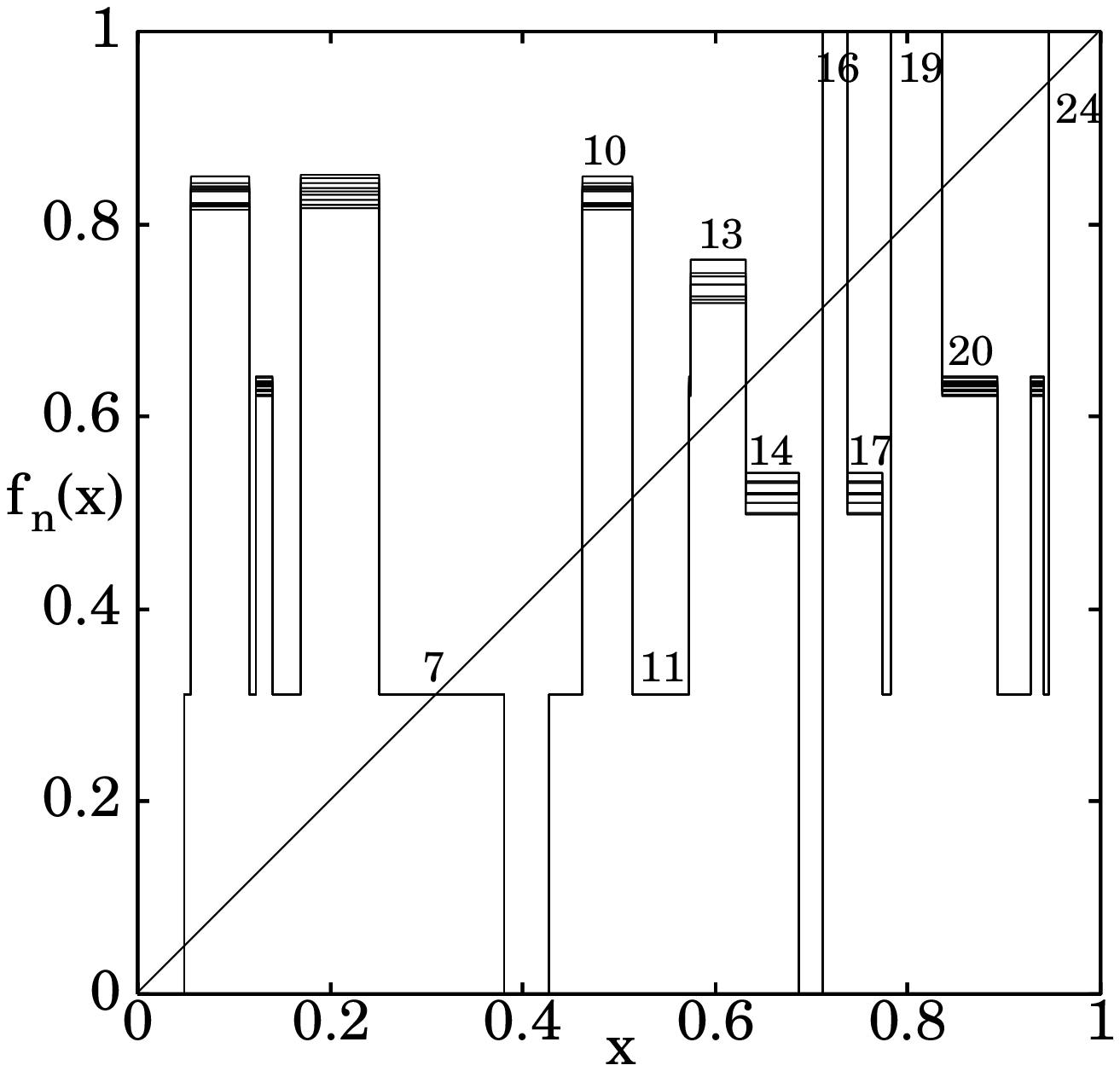} &
\includegraphics[width=6cm,height=6cm]{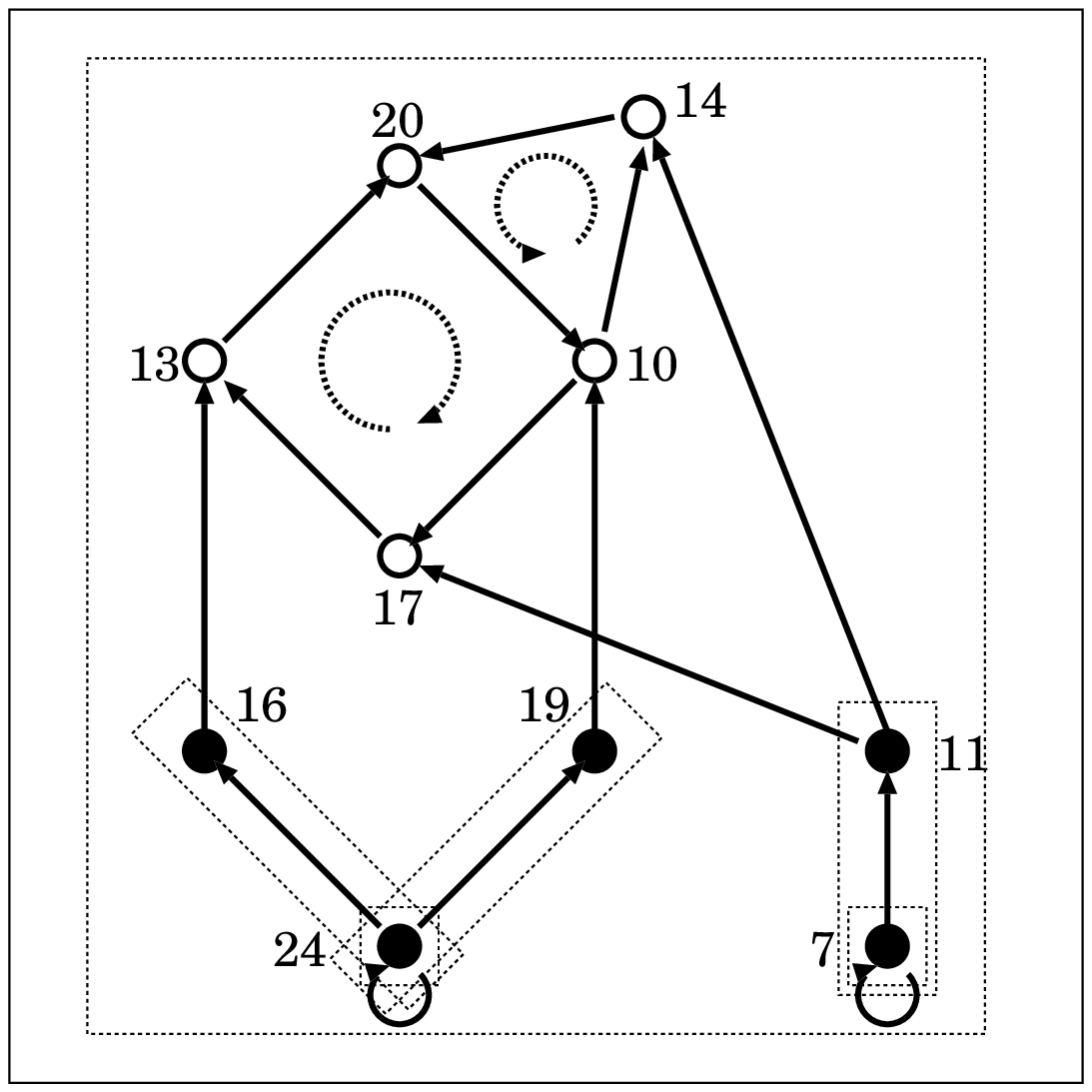}\\
\end{tabular}
\caption{\label{fig:random}{\small Two typical results are displayed in (a) and (b)
for tha case $\e = 0.1$. Each partial function plotted is on the attractor.
In both cases, the simulation started from a random piecewise constant $f_0$.
In ($\cdot$-1), the graphs of $f_n$ $n = 1000$--$1010$ are overlaid.
In ($\cdot$-2), the corresponding network is displayed. 
The index $i$ with a filled circle indicates
that $f_n|_{I^i}$ is $n$-independent, 
while that with an unfilled circle indicates that $f_n|_{I^i}$ is $n$-dependent.
A square drawn with a broken line indicates a set of indices 
$\lambda$ that satisfies $S(\lambda)\subseteq \lambda$.
(a-1) The period of $f_n|_{I^6}$ is 5 and the period of $f_n|_{I^2}$ and $f_n|_{I^{13}}$ is 10.
(b-2) The period of all periodic partial functions is 145
(although the phase of the oscillation is different for each subinterval).
}
}
\end{center}
\end{figure}

Using $f_0$ as defined in the previous subsection,
we numerically computed the temporal evolution of $f_n$.
In Fig.\ref{fig:random}, two typical results (excluding transients)
are displayed.
In this set of simulations, we choose $N = 30$ and $\e = 0.1$.
We find that some neighboring subintervals are mapped to the same value
after the transient behavior has ended, i.e.,
$f_n(I^i) = f_n(I^{i+1})$ for sufficiently large $n$.  
In this case, these subintervals join to form a new subinterval.
In such situation, the indices of the subintervals are renumbered.
[Hence the maximal value of the index $i$ for the subintervals $I^i$ can become smaller than $N-1$ 
(26 in Fig.\ref{fig:random}(a) and 24 in  Fig.\ref{fig:random}(b)).]

For most initial functions, $f_n(x)$ falls on an attractor rapidly (by $n\sim 100$).
In Figs.\ref{fig:random}($\cdot$-1), the graphs of $f_n$ ($n = 1000$--$1010$) are overlaid.
As three typical examples we show the behavior of $f_n|_{I^6}$, $f_n|_{I^2}$
and $f_n|_{I^{13}}$ specifically. In (a-1), the period of $f_n|_{I^6}$ is 5 and the period of $f_n|_{I^2}$ and $f_n|_{I^{13}}$ is 10.
In (b-1), the period of all periodic partial functions is 145.
Although there seems to be no easily discernible difference between two time series,
there is a significant difference between the structures of their networks.

We now set out to describe the function dynamics in terms of network dynamics.
First, let $S_n(i)$ be a function mapping a set of indices to itself.
We define $S_n(i)$ and $S(i)$ as follows: 
$S_n(i) = j$ if $f_n(I^i) \subset I^j$ and 
$S(i) := \bigcup_n S_{n+m}(i)$ for sufficiently large $n$ and $m$. 
In other words $S(i)$ is the set of all $S_n(i)$ after transient behavior.
A directed edge $a\rightarrow b$ in Figs.\ref{fig:random}($\cdot$-2) is drawn if 
$a \in S(b)$.\footnote{The direction of the edge is opposite to the direction of the mapping $f_n$.
Here the direction of the edge is chosen to be in the direction of the 
propagation of a perturbation applied to at one partial function, which is discussed
in the next subsection.}

Now, let $\lambda$ be a set of indices. Here,
two classes of $\lambda$ are important for later discussion.
\begin{itemize}
\item i) The class for which $S(\lambda)\subseteq \lambda$. 
In this case, $f_n|_{I^i}$ ($i \in \lambda$) evolves independently.
This means that 
the time evolution of $f_n|_{I^i}$ ($i \in \lambda$) is not influenced by a
perturbation applied to the value of $f_n|_{I^j}$ for any $j \not\in \lambda$.

If two sets of indices $\lambda_0$ and $\lambda_1$ satisfy this condition 
and if $\lambda_0 \subset \lambda_1$,
we say that `$\lambda_0$  and $\lambda_1$' are {\em hierarchical}.
\footnote{We call the generated map $g_n|_{I^i}$ ($i\in \lambda_1$) a `meta-map' \cite{NKKII}.}
In this situation, a perturbation applied to $f_n|_{I^i}$ ($i \in \lambda_0)$ can 
influence the evolution of  $f_n|_{I^j}$ ($j \in \lambda_1\setminus \lambda_0$),
but a perturbation applied to $f_n|_{I^j}$ ($j \in \lambda_1\setminus \lambda_0$) does not
influence the evolution of $f_n|_{I^i}$ ($i \in \lambda_0$).

\item ii) The class for which $S(\lambda) = \lambda$. In this case, if $\lambda$ consists of
a single element, it is a fixed point.
If $\lambda$ consists of more than one elements
the network formed by $S(\lambda)$ possesses a loop. This means that 
$i \in S(i) \cup S(S(i)) \cup S(S(S(i)))\cup \cdots$ for some indices $i$.
Then, if some perturbation is applied to at $f_n|_{I^i}$ for $i$ contained in this loop, 
the effect propagates through the loop and eventually returns to $i$ itself.
In this case, we say that $\lambda$ is `{\em entangled}'.
\end{itemize}

In Figs.\ref{fig:random}($\cdot$-2), 
sets $\lambda$ satisfying $S(\lambda)\subseteq \lambda$ are contained within  squares drawn 
with broken lines, while entanglements are
indicated  by loops drawn with broken lines.
As shown, the network in Fig.\ref{fig:random}(a) is hierarchical with four layers,
and there is no entanglement.  
Contrastingly, the network in Fig.\ref{fig:random}(b) has two entanglements (two loops).

The shape of the network is determined by the 
manner in which intervals are related under $f_n$,
and the shape can be preserved even if  the period of each partial function
is changed by changing the coupling $\e$. 
Recall that the shape of the network is resulting from the contracting nature of the generated map.
For this reason the shape can be preserved when a small change is made to
the parameter $\e$, and
the network is stable with respect to the changes in $\e$.
Also, a network can be stable with respect to noise applied to each
partial function.
In the next subsection, we study the effect of noise on the network.

\subsection{Entangled Network}

Dynamical networks in FD systems can possess hierarchy and entanglement.
As discussed above one difference between a network with entanglement and one without entanglement
regards the behavior resulting from perturbation.
In this subsection, we study the effect of noise
on dynamical networks.

If a network is hierarchical and contains no loops (no entanglements),
the evolution of $f_n$ can be ordered unidirectionally: 
$n$-independent partial functions determine an $n$-independent generated map, which drives $n$-dependent partial functions,
which determine an $n$-dependent generated map,
which drives other $n$-dependent partial functions, and so forth.\footnote{
In Ref.\cite{NKKII}, this hierarchy is represented by meta-meta-... maps.}
In this case, the hierarchy of relations among partial functions 
is uniquely ordered through generated maps.

Now, we consider how a perturbation applied to one partial function
propagates.  For this purpose, we apply ``noise" to one partial function,
changing $f_n|_A$ to $f_n|_A + \delta 1^A$, with $\delta$ as a small
number.  Here
the noise is added only at the $n$-th step, and later time
evolution of the function is given by (1) without noise.
We now study how
the effect of such a perturbation is transmitted through the network.

In the case of a hierarchical network without entanglement,
this noise propagates in one direction.  In this case,
there are three types of noise effects: 

\noindent i)   Noise destroys the network, and a new network appears.

\noindent ii)  The effect of the noise disappears due to the contraction induced by the generated map.

\noindent iii) The effect of noise propagates unidirectionally the network, while 
the shape of the network is preserved.  
In this case, only the phase of oscillation
of each partial function $f_n|_A$ is changed.

If the noise amplitude $\delta$ is not too large, the network shape
is not destroyed, and the only important effect is that of type (iii).
In this situation, the phase change of a partial function causes 
a phase change of the corresponding $n$-dependent generated map.
As a result, the phase of the partial function driven by this generated map changes,
and so forth.
Hence, the noise effect for a sufficiently small amplitude $\delta$ 
in such a hierarchical network consists of 
represented as unidirectional propagation of phase changes
through the network.

In the case of a network with entanglement, 
another type of noise effect, which is caused by loops, appears:

\noindent (iv) The noise effect circulates through a loop and leads to
a transition to a new attractor.

In this case, the noise propagates unidirectionally in the loop
and returns to the element to which the perturbation was applied. 
This returned noise changes the generated map again, 
and this effect, if it does not decay away, returns to the same element again.
Due to the continuation of this process, a new attractor on the loop is formed.
This behavior is in strong contrast with that seen in
hierarchical networks without loops, in which only the phase of
oscillation is changed.

\subsubsection{Simple Entanglement: An example}

\begin{figure}[htbp]
\begin{center}
\begin{tabular}{ll}
(a) & (b)\\
\includegraphics[width=6cm,height=6cm]{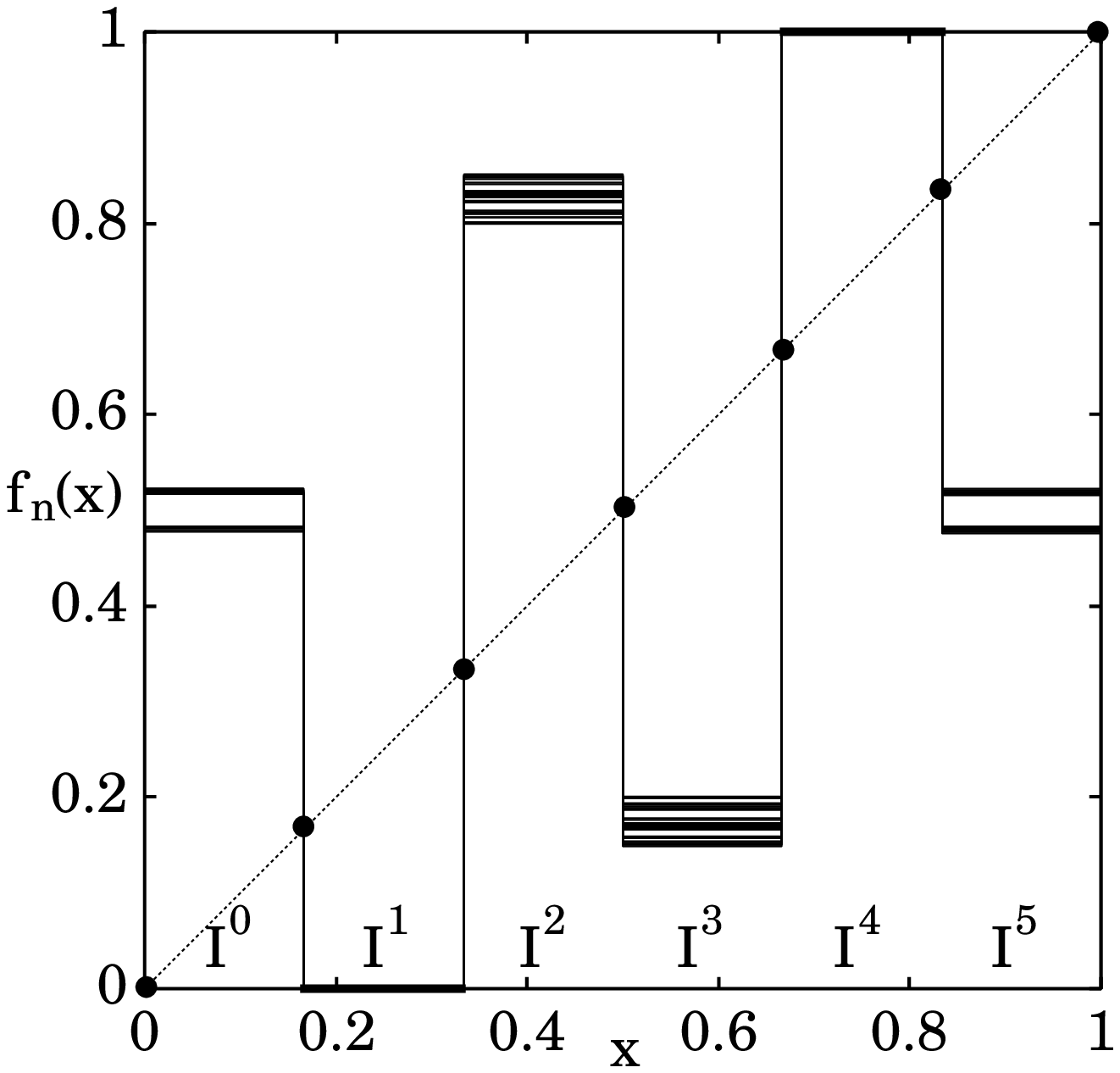}&
\includegraphics[width=6cm,height=6cm]{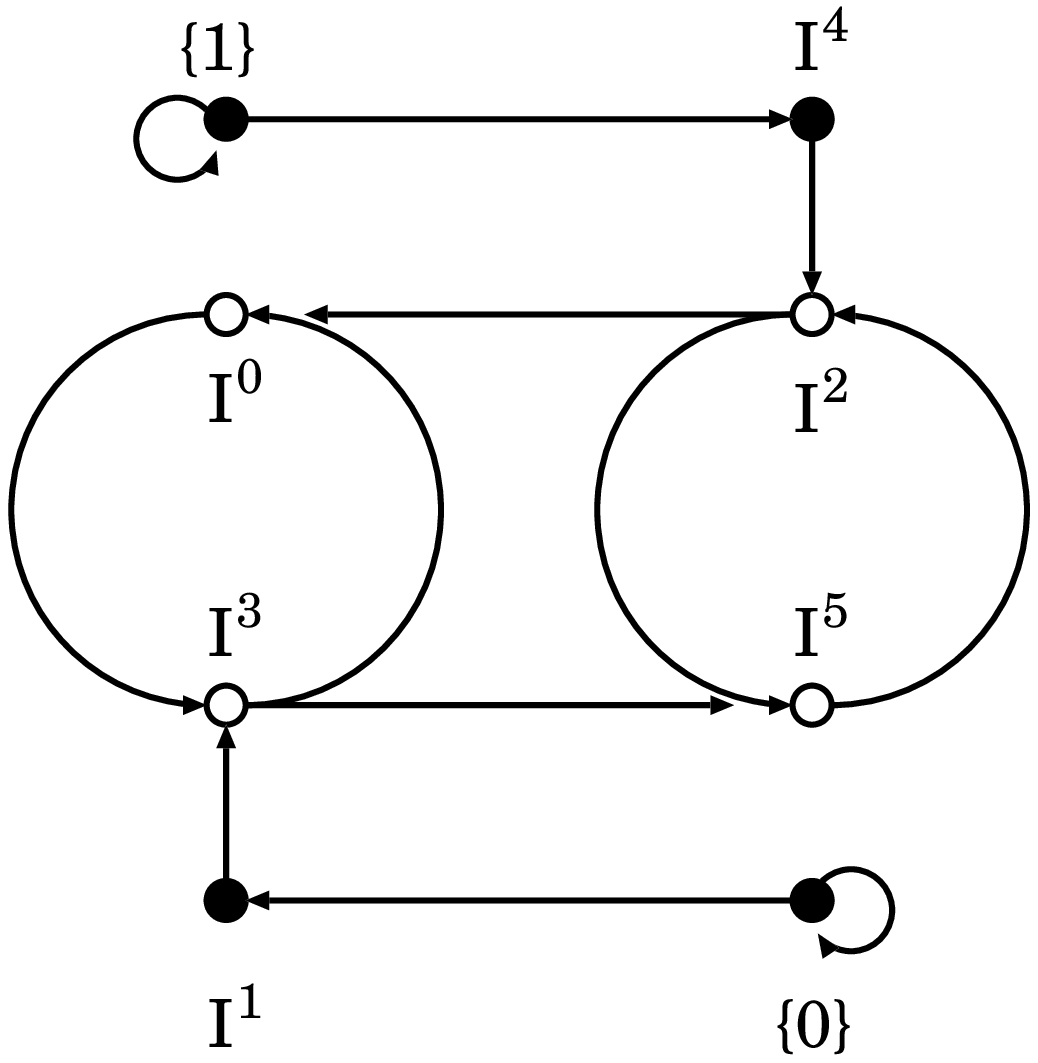}\\
\end{tabular}
\caption{\label{fig:entangle}{\small (a) The graph of a limit function $f_n$ (for $n = 1000$--$1026$).
Here, ${a^0}_0 = 1/3 + 1/20$, ${a^2}_0 = 2/3 + 1/10$, ${a^3}_0 = 1/10$, ${a^5}_0 = 1/3 + 1/10$ and $\e = 0.11$.
The period is 26.
(b) The network corresponding to this function.
There are three loops.
}
}
\end{center}
\end{figure}

Now, we study the dynamics of a network with entanglement,
using several different initial functions.
First, let $I = [0, 1]$ be divided into the subintervals 
$I^i = (i/6, (i+1)/6)$ $(i = 0, \ldots, 5)$, and the points
$\{i/6\}$ ($i = 0, \ldots, 6$).
(Obviously, $\displaystyle I = (\bigcup_0^5 (I^i))\cup (\bigcup_0^6 \{i/6\}.)$
We define the initial function $f_0$ according to Eq.(4), 
in terms of the parameter ${a^i}_0$.

\begin{equation}\label{eq:5}
\left\{
\begin{array}{ll}
f_0(\{i/6\}) & = i/6\\
{a^0}_0    & \in I^2\cup I^3\\ 
{a^1}_0    & = 0\\
{a^2}_0    & \in I^4\cup I^5\\
{a^3}_0    & \in I^0\cup I^1\\
{a^4}_0    & =1\\
{a^5}_0    & \in I^2\cup I^3\\
\end{array}
\right.
\end{equation}

In Fig.\ref{fig:entangle}(a), an example of the limit function 
resulting from such an $f_0$ is displayed. In that case, 
the set of initial values ${a^i}_0$ were chosen as
${a^0}_0 = 1/3 + 1/20$, ${a^2}_0 = 2/3 + 1/10$, ${a^3}_0 = 1/10$, 
${a^5}_0 = 1/3 + 1/10$ and $\e = 0.11$ was used.
In the figure, the graphs of $f_n$ for $n = 1000$--$1026$ are overlaid.
The period of $f_n$ here is 26.

\begin{figure}[htbp]
\begin{center}
\includegraphics[width=12cm,height=15cm]{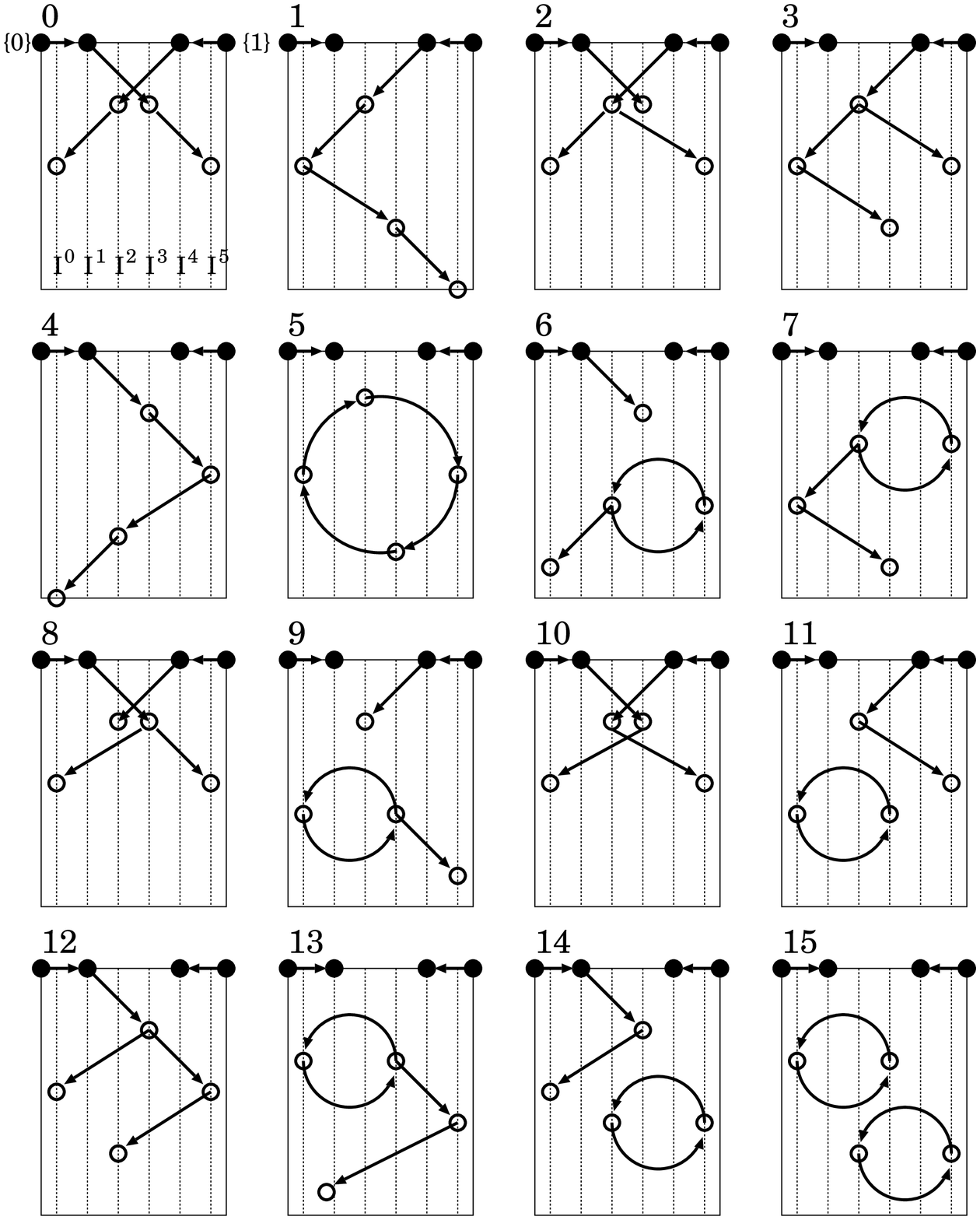}
\caption{\label{fig:16}{\small The sixteen possible states of the network at 
any given time step $n$ allowed by the initial nature of $f_0$ as defined in Eq.(5).
}
}
\end{center}
\end{figure}

The network for this function is shown in Fig.\ref{fig:entangle}(b).
As seen, there are three loops.
It was found that if $\e < 1/6$, the shape of the network does not change from the initial time, 
because the generated map satisfies the condition 
$g_n|_{I^{2i}\cup I^{2i+1}} \subset I^{2i}\cup I^{2i+1}$ for $i = 0, 1, 2$.
Due to the nature of $f_0$ given by Eq.(\ref{eq:5}), the possible types of network structure
are restricted.  
First, the partial functions $f_n|_{I^1}$ and $f_n|_{I^4}$ are $n$-independent.
The other four $f_n|_{I^i}$ are $n$-dependent and refer to some $I^{2j} \cup I^{2j+1}$
(for example, $f_n(I^0) \in I^2$ or $I^3$).
Thus each of the four $n$-dependent partial functions refers to either of 
two subintervals at each time step.
Accordingly, there are $2\times 2\times 2\times 2$ possibilities at each step.
In Fig.\ref{fig:16}, the sixteen possible networks 
that are allowed for this type of initial function are displayed as
the directed edges $S_n(i) \rightarrow i$.
At each time step, $f_n$ corresponds to one of these sixteen possible networks.
We call the number in the figure the `state' of the network.
Since the number of states, 16, is smaller than the period, 26, 
the network must exist in the same state (but with different values
of the partial functions) at different times in a single period during the evolution.

\begin{figure}[htbp]
\begin{center}
\begin{tabular}{ll}
(a-1) & (a-2)\\
\includegraphics[width=6cm,height=6cm]{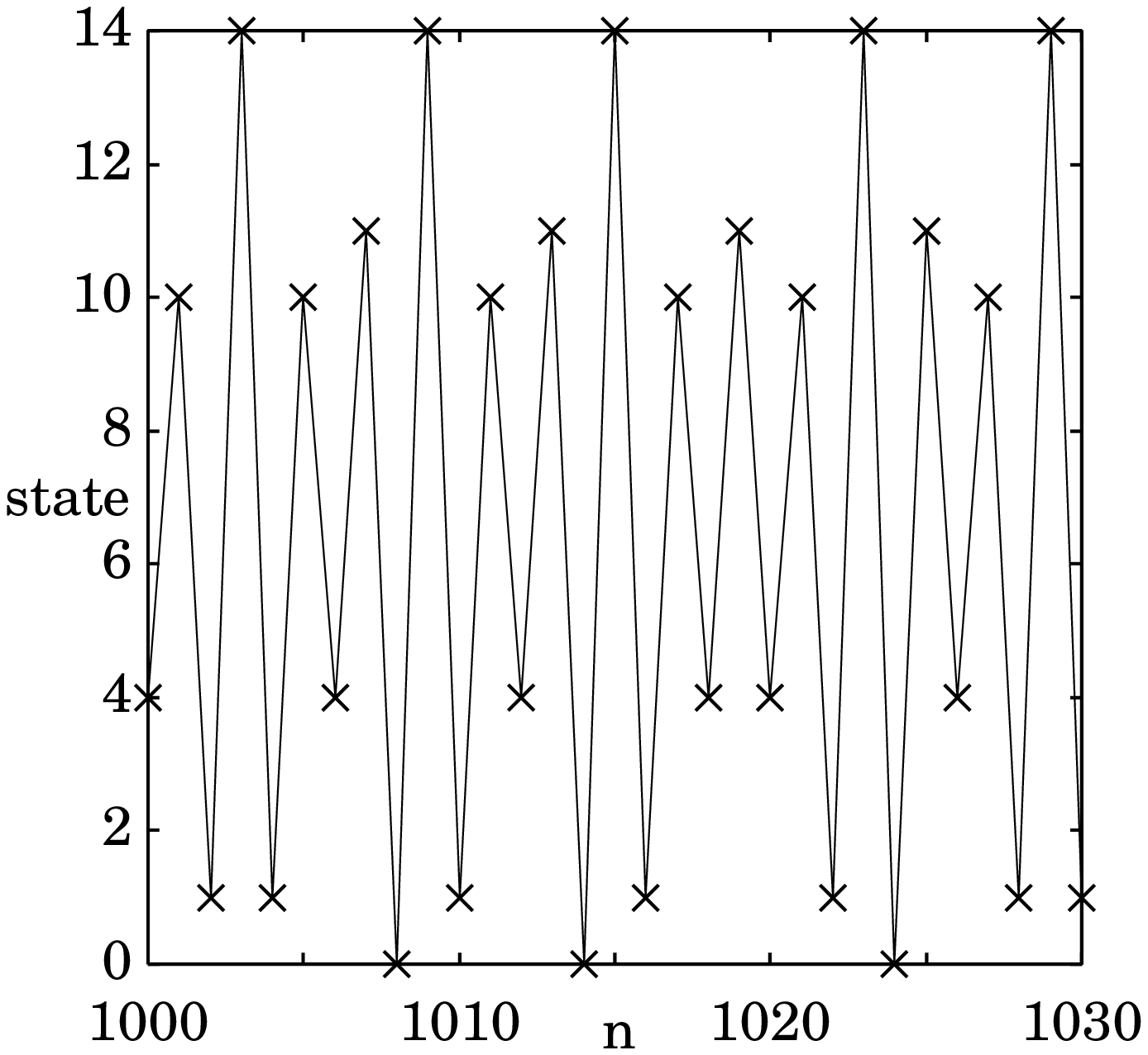}&
\includegraphics[width=6cm,height=6cm]{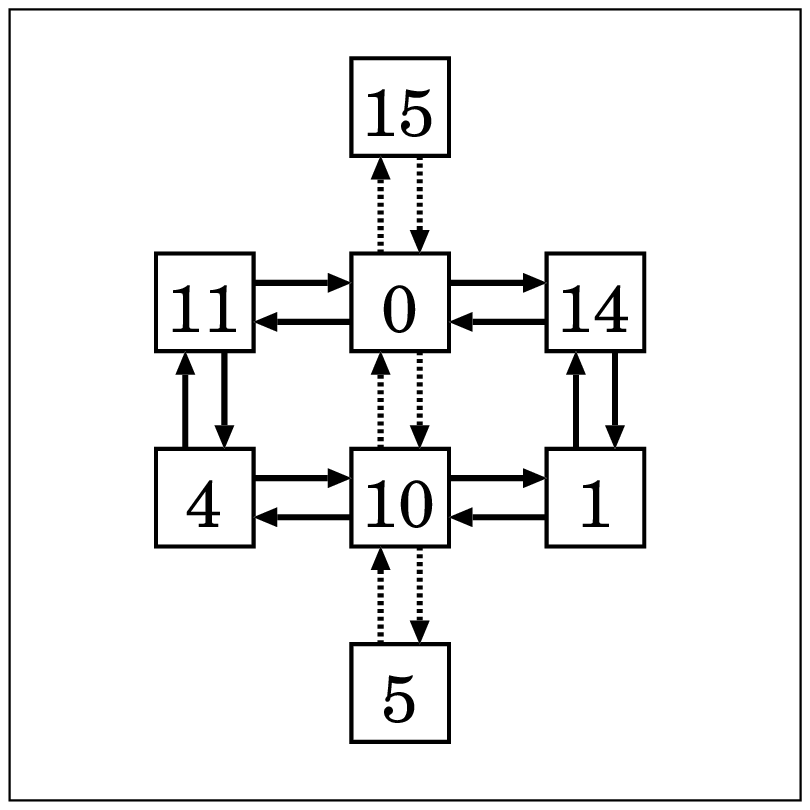}\\
(b-1) & (b-2)\\
\includegraphics[width=6cm,height=6cm]{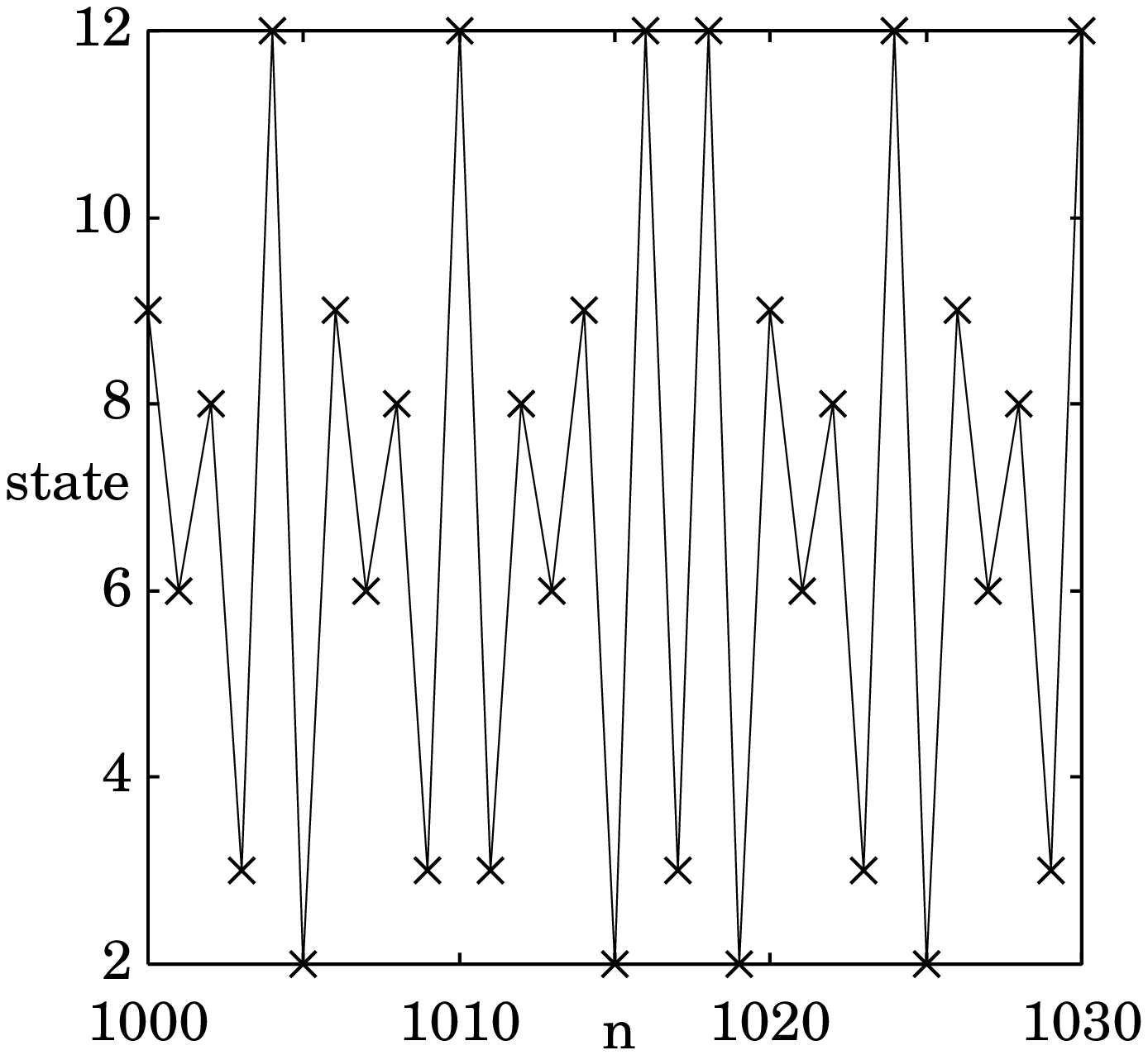}&
\includegraphics[width=6cm,height=6cm]{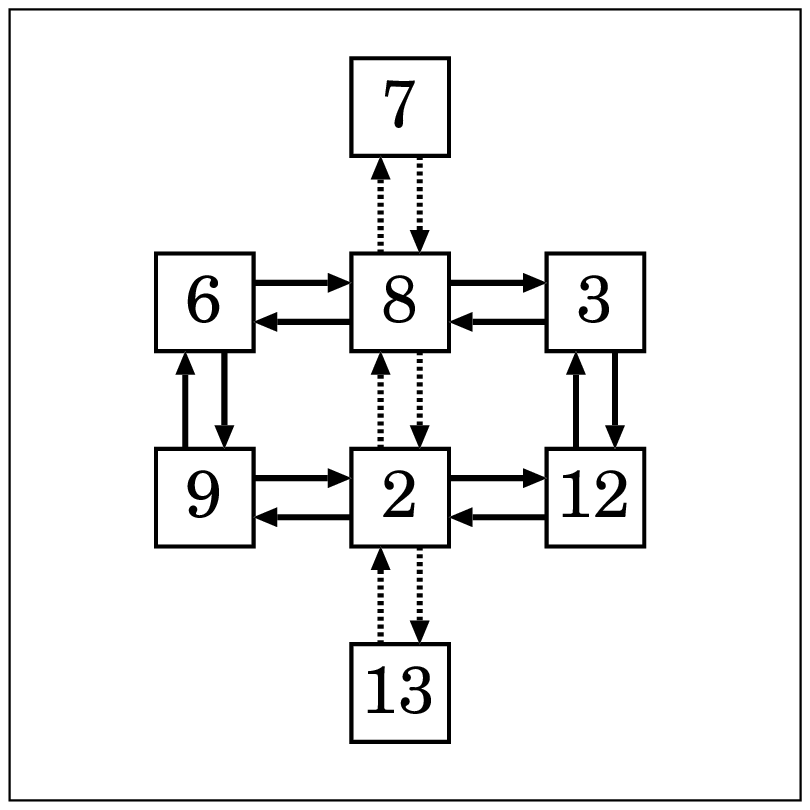}\\
\end{tabular}
\caption{\label{fig:state}{\small In (a-2) and (b-2) are examples 
of the sequence of transitions between states for the FD system with $\e = 0.11$.
The initial function in (a-1) is defined by Eq.(5) with 
${a^0}_0 = 1/3 + 1/20$, ${a^2}_0 = 2/3 + 1/10$, ${a^3}_0 = 1/10$, ${a^5}_0 = 1/3 + 1/10$, 
and [note that $f_0(I^0) \neq f_0(I^5)$],
and the initial function in (b-2) is defined by Eq.(5) with 
${a^0}_0 = 1/3 + 1/10$, ${a^2}_0 = 2/3 + 1/10$, ${a^3}_0 = 1/10$, ${a^5}_0 = 1/3 + 1/10$ 
[note that $f_0(I^0) = f_0(I^5)$].
(a-1) and (b-1) depict the evolution of the state from $n=1000$ to $n=1030$.
The period of the sequence of transitions depicted in (a-1) is 26, 
and that of the sequence depicted in (b-1) is 20.
In (a-2) and (b-2), the possible sets of transitions are depicted.
Here (a-2) corresponds to the case $f_0(I^0) \neq f_0(I^5)$,
and (b-2) corresponds to the case $f_0(I^0) = f_0(I^5)$.
These two figures depict the two types of transition sets possible in general, 
while the actual transitions possible for any given choices of $f_0$ 
will consist of some subset of these.
For the particular choices considered here,
only the transitions denoted by solid arrows occur.
}
}
\end{center}
\end{figure}

The FD system we study can be described in terms of the dynamics of a network,
consisting of successive transitions between the states defined in the above manner.
The actual set of allowed transitions between states depends on
the initial values ${a^i}_0$.
By examining the FD of each partial function,
case by case, one can show that
possible sequences of transitions are of two types, as defined in 
Figs.\ref{fig:state}(a-2) and \ref{fig:state}(b-2).
These two types correspond to the cases 
in which $f_0(I^0) \neq f_0(I^5)$ and $f_0(I^0) = f_0(I^5)$.  
There are two types of possible sets of transitions exhibited by the FD system,
depicted schematically by Figs.\ref{fig:state}(a-2) and (b-2).
Each type includes the transitions denoted by both solid and broken arrows in the figures,
but for given ${a^i}_0$, the actual transition sequence includes only some of the transitions indicated.
For the particular choices of $f_0$ considered here, only the transitions denoted by solid arrows occur.

Two examples of the sequence of transitions between states 
starting from two different initial functions are plotted in Fig.8.
Here, two actual sequences of transitions between states from $n = 1000$ to $n = 1030$ are 
plotted in Figs.\ref{fig:state}(a-1) and (b-1), where
$f_0(I^0) = f_0(I^5)$ is satisfied in (b-1) and not in (a-1).
As shown in Fig.\ref{fig:state},  
the set of possible transitions depends on 
whether $f_0(I^0) = f_0(I^5)$ is satisfied or not.
This set in the former case is depicted in (b-2) and in the latter case in (a-2).
(Note that these sets of transitions characterize long-time behavior, 
i.e., after transient behavior has ended.)

Now, the function $f_n$ on an attractor is characterized on three levels,
according to i)   the value of $f_n|_{I^i}$,
ii)  the state of the network, and 
iii) the set of possible transitions given in Figs.\ref{fig:state}($\cdot$ -2).

\subsubsection{Transitions among Attractors in Entangled Networks}

\begin{figure}[htbp]
\begin{center}
\includegraphics[width=8cm,height=8cm]{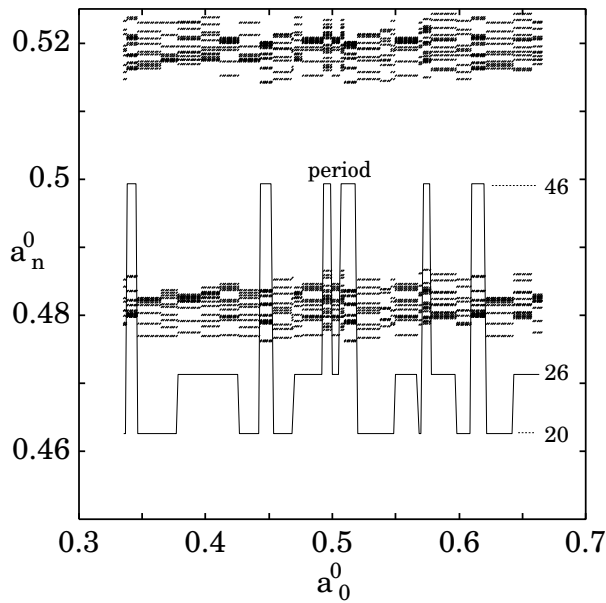}\\
\end{center}
\caption{\label{fig:basin}{\small The limiting values of ${a^0}_n$ respect to ${a^0}_0$.
Here, ${a^2}_0 = 2/3 + 1/10$, ${a^3}_0 = 1/10$, ${a^5}_0 = 1/3 + 1/10$ and $\e = 0.11$.
The period of ${a^0}_n$ (which is also the period of $f_n$) is plotted by the solid curve.
}
}
\end{figure}

As discussed above, the behavior of the network dynamics in a FD system depends on
the initial conditions, represented by the 
values ${a^i}_0$.  
Although the parameter characterizing the FD model, $\e$, is fixed,
the change undergone by the FD attractor as the ${a^i}_0$ are changed is
bifurcation-like, because the generated map changes
with the values of the ${a^i}_0$.
To elucidate the dependence on the initial conditions,
we carried out numerical simulations using the definition given in Eq.(5),
employing various values of ${a^0}_0$, with all other ${a^i}_0$ ($i = 1, \ldots 5$)
values fixed.  
In Fig.\ref{fig:basin}, the values 
${a^0}_n$ on an attractor (after transients) are plotted 
with respect to ${a^0}_0$.
The solid curve represents the period of the attractor of $f_n$ 
resulting from the corresponding initial value ${a^0}_0$.
The flat pieces of this curve express the meaning that 
over the corresponding interval of ${a^0}_0$ values ${a^0}_n$ is on the same attractor.
The stability of attractors with respect to changes of ${a^0}_0$ 
that the existence of these flat pieces reflects  
arises from the property that the slope of the generated map is less than unity.
The discontinuities of this curve are caused by 
discontinuities of $f_{\infty}$.

\begin{figure}[htbp]
\begin{center}
\begin{tabular}{ll}
(a) & (b)\\
\includegraphics[width=6cm,height=6cm]{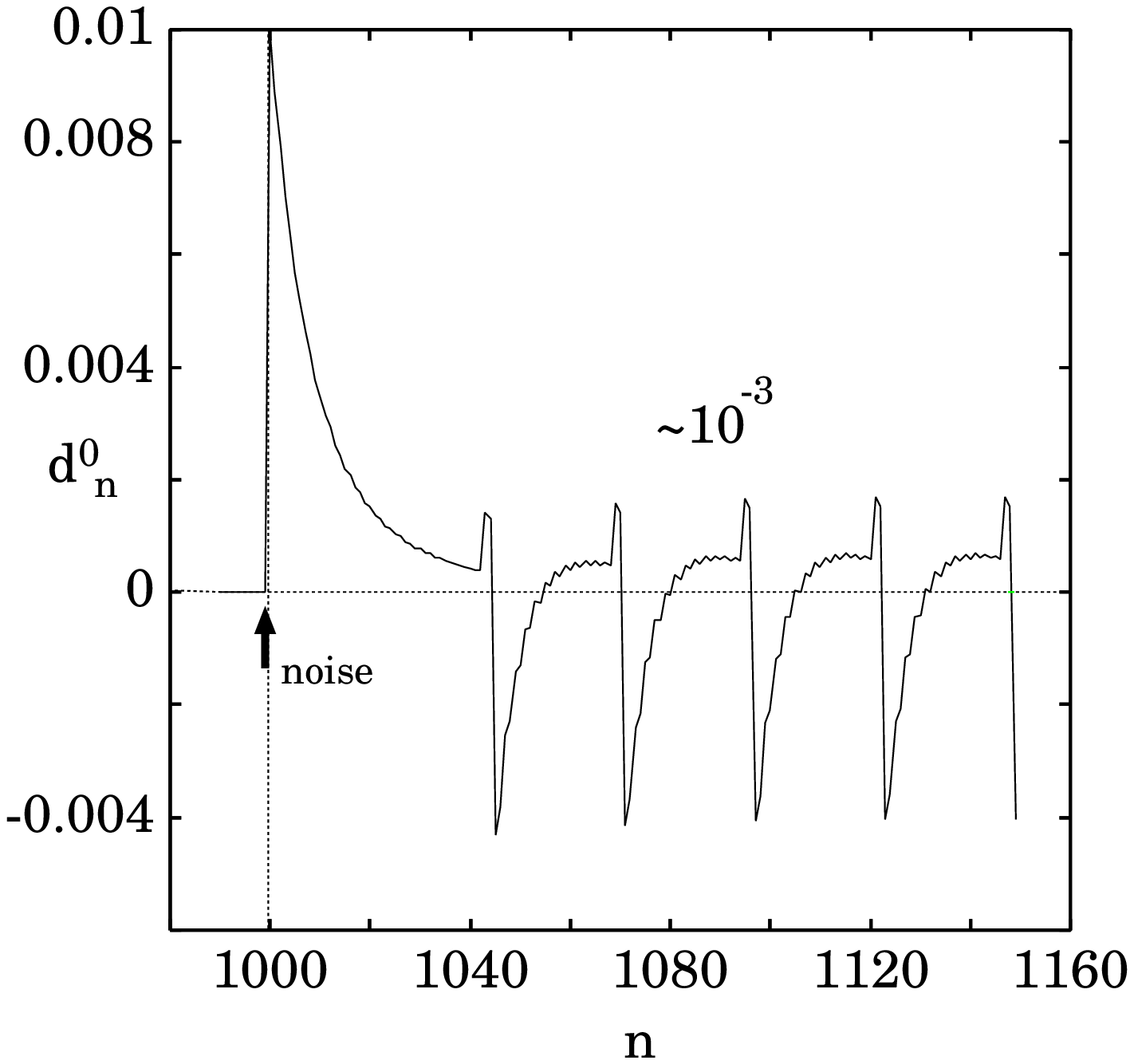}& 
\includegraphics[width=6cm,height=6cm]{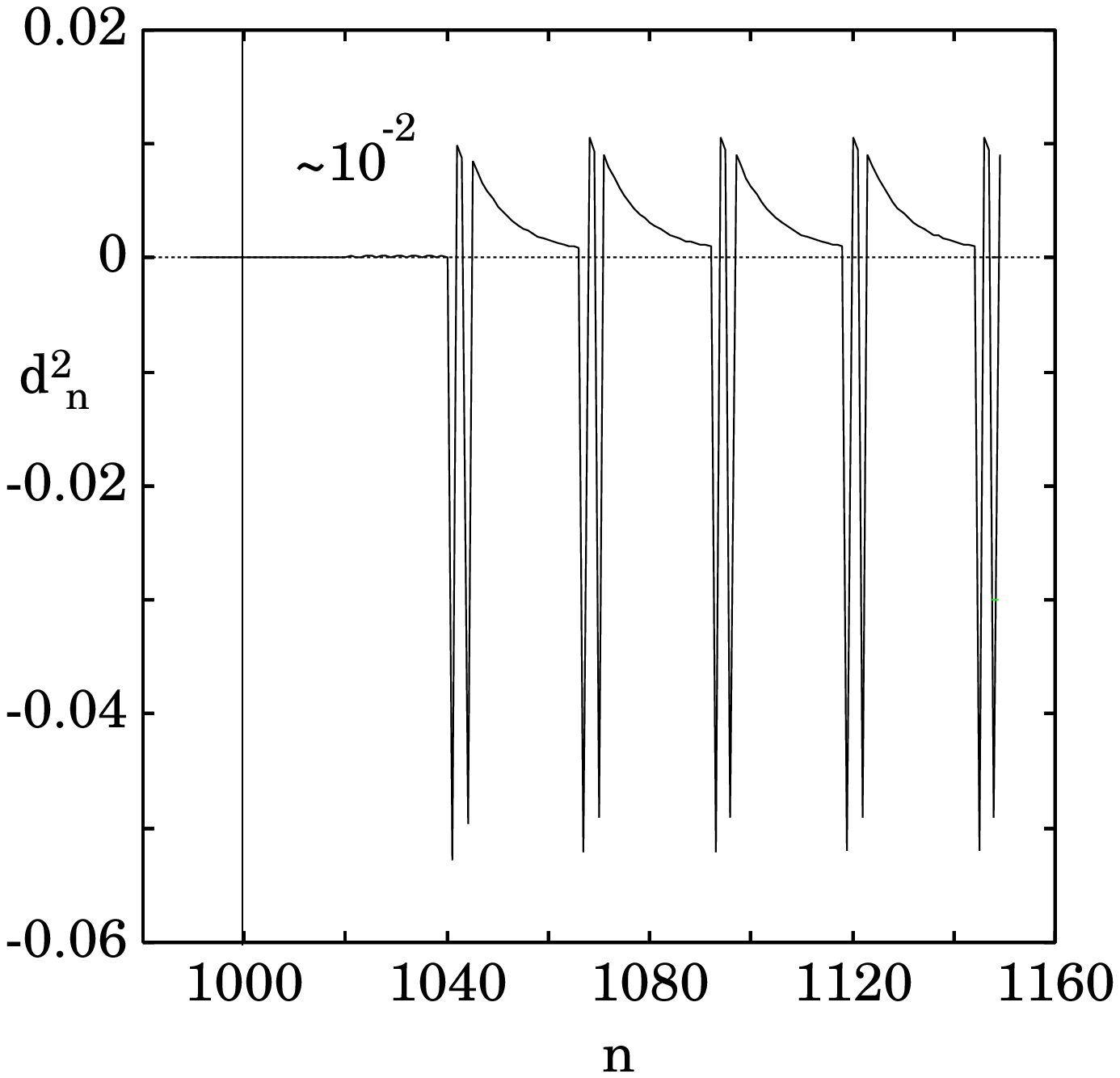}\\

(c) &(d)\\
\includegraphics[width=6cm,height=6cm]{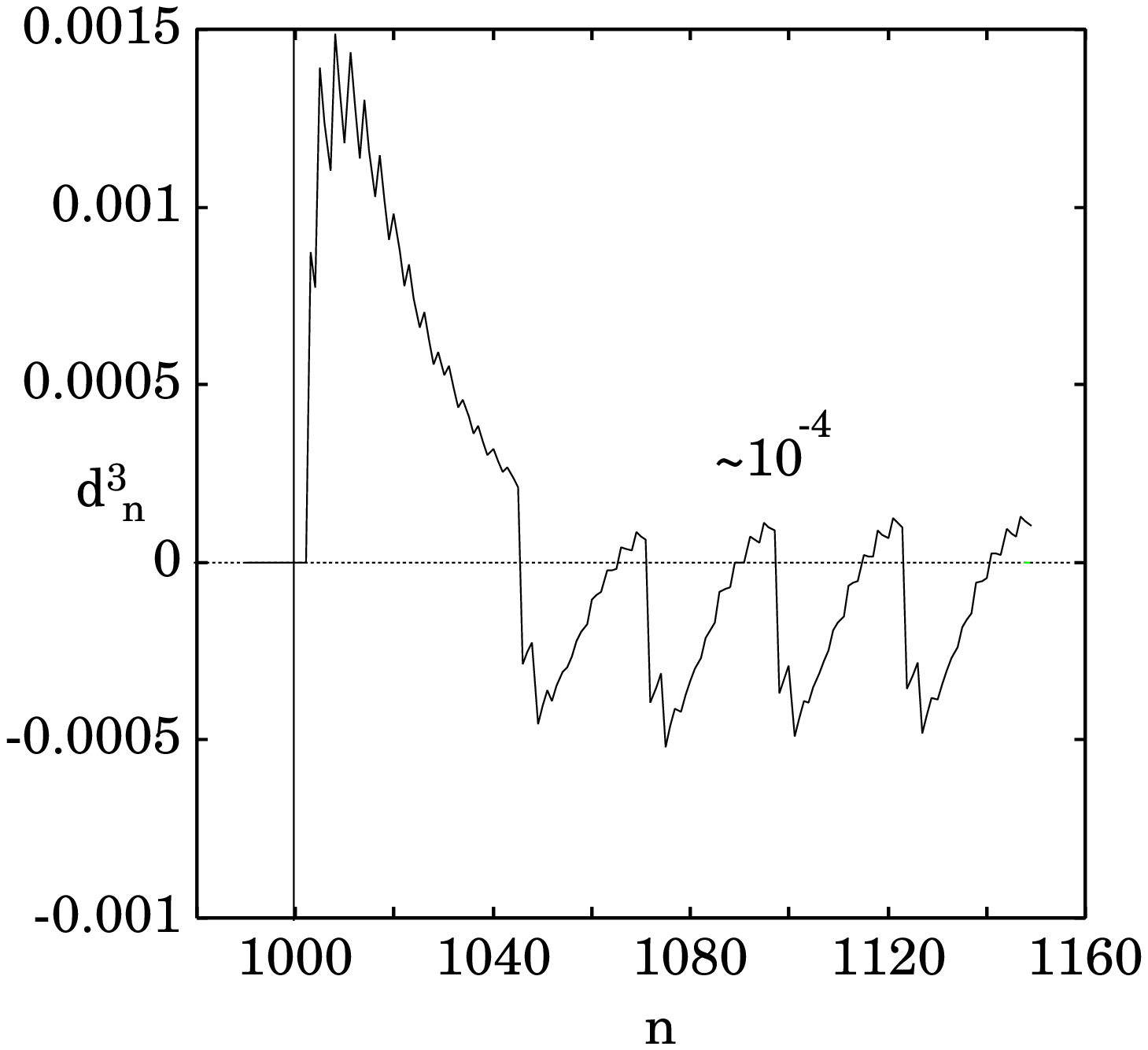}& 
\includegraphics[width=6cm,height=6cm]{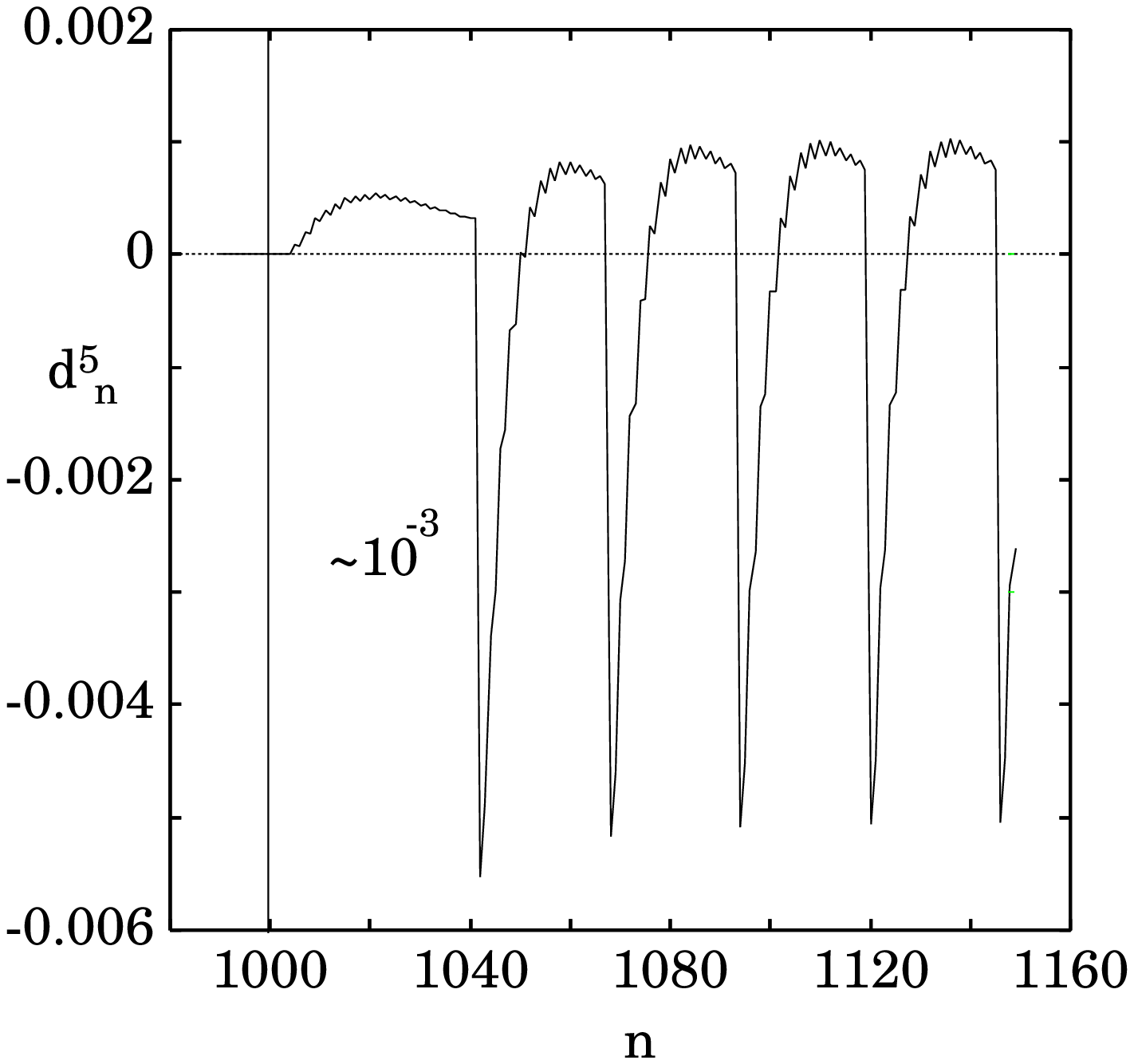}\\
\end{tabular}
\end{center}
\caption{\label{fig:noise}{\small The time series of the difference, $d_n$, between the  original 
function $f_n$ and $h_n$, obtained by perturbing $f_n|_{I^0}$ at $n=1000$.
(a)${d^0}_n$, (b)${d^2}_n$, (c)${d^3}_n$, (d)${d^5}_n$.
The initial function used here is defined by Eq.(5), with 
${a^0}_0 = 1/3 + 1/20$, ${a^2}_0 = 2/3 + 1/10$, ${a^3}_0 = 1/10$, 
${a^5}_0 = 1/3 + 1/10$. We also used $\e = 0.11$.
Note that the scales of the vertical axes differ among the figures.
}
}
\end{figure}

By applying a perturbation to an attractor, a transition
between attractors can result.  
Such a transition, on the other hand, can be understood as a 
propagation of noise through the network, as mentioned above. 
To see how such noise can propagate, we applied an instantaneous ``noise'' to
$f_{n}|_{I^0}$, for sufficiently large $n$ that $f_n$ is on an attractor. 
(We took $n=1000$ avoid transients.)
We computed the time evolution of the function after this perturbation,
which is denoted $h_n$, and then measures the
difference ${d^i}_n$ between $h_n$ and the unperturbed $f_n$:
${d^i}_n=h_n|_{I^i} - f_n|_{I^i}$.
In Fig.\ref{fig:noise}, time series of ${d^i}_n$ are plotted.

The transitions among attractors progress as follows.
At $n = 1000$, the instantaneous noise is added.
This noise first is transmitted to $f_n|_{I^3}$ (see Fig.\ref{fig:entangle}(c)).
However, the effect of the perturbation 
decays on $I^0$ and $I^3$, as seen from $d^0_n$ plotted in (a) and $d^3_n$ plotted in (d)
of Fig.\ref{fig:noise}.
Then, $f_n|_{I^5}$ is influenced by the noise.  Through $I^5$, the noise 
effect is conveyed to $f_n|_{I^2}$, as seen in (b).
For $n\sim 1040$, the value of $f_n|_{I^2}$ changes and this
effect is transmitted back to $f_n|_{I^0}$.  Hence, the noise effect circulates 
between $I^0$ and $I^2$, and
for $n > 1040$, $f_n|_{I^2}$ acts as a `pulse generator'.
Here, the amplitude of the `pulse' is of the same order as the noise.
Note that the magnitude of ${d^i}_n$ is of a different order on each $I^i$:
(a) ${d^0}_n\sim 10^{-3}$, (b) ${d^2}_n\sim 10^{-2}$, 
(c) ${d^3}_n\sim 10^{-4}$, (d) ${d^5}_n\sim 10^{-3}$.
The noise effect is amplified on $I^5$ 
and is further amplified on $I^2$.
From this point of view, we can consider the added noise to be stored in $f_n|_{I^2}$, 
which leads to a transition to a new attractor.
This transition is possible because the network is entangled.
In this case the noise effect returns to the interval where the noise was originally added.
Then, deviation from the original attractor is sustained within the loop
in which the noise effect is circulated.  Hence the function switches
to a new attractor.

\subsection{Memory in a Dynamical Network}

As is already discussed in Ref.\cite{NKKI}, our FD can be regarded as a system 
that transforms input to output through $f_n(x)$, with an
input-output relationship that changes autonomously.
Since such an autonomous change of an input-output relationship is
commonly observed in  biological cognitive systems,
it is interesting to consider the possible function of an entangled network
in this context.

When our function $f_n$ is regarded as defining an input-output relationship,
the $a^n_0$-$a^0_0$ relation in Fig.\ref{fig:noise} represents 
the input dependence of the output.  
This relationship exhibits several thresholds with regard to change in the output behavior;
when the input is changed beyond such a threshold, the output switches
to new behavior.
This switch can be caused by perturbing a partial function.
The pulse generated by such a perturbation can be regarded as a memory.
Within a network with loops, the input (a perturbation applied to one
partial function at one time step) is memorized through the propagation of noise
within the loop.
Note that for this type of network, i) the attractor can be changed as a result of this noise through a
transition process and that ii) information can be stored in some 
parts of the network by virtue of the circulation of the input.
It would thus appear that a dynamical network of this kind has the potentiality to 
possess memory that depends on an input.

\section{Summary and Discussion}

\subsection{Summary}

In the present paper, FD has been studied by constructing 
initial functions suitable for analytical and numerical study,
after basic introduction to a FD model in Sec.2.

It was shown that the dynamics of a FD system are represented by a generated map, 
a 1-dimensional map $g_n$ determined by the 1-dimensional map $f_n$.
The generated map acts as the rule governing the temporal evolution of
$f_n(x)$ (i.e., $f_{n+1}(x)$ is determined by $g_n$).
For most initial conditions, after transient behavior dies away,
the function $f_n$ is attracted to a piecewise-constant function.
In other words, $f_n$ comes to be divided into partial functions, 
each of which is constant.  In Sec.3, the stability of these
piecewise-constant functions was discussed.  
In Sec.4 we considered the behavior resulting when we
start from a piecewise-constant function, where each constant partial
function is regarded as an element of a network.  The dynamics of
such a network, which generate a rule to drive the network itself,
were also studied in Sec.4.

After defining initial functions appropriate to allow for the behavior described above in Sec.4.1,
the dynamics of the resulting type of network were classified in Sec.4.2.
In one class, the dynamics of the partial functions are
determined hierarchically. These partial functions are unidirectionally ordered
in a tree, in which each partial function in an upper level is driven
by partial functions by lower levels.
In this case, the effect of a perturbation  applied to a partial
function is transmitted unidirectionally through the 
partial functions.

In Sec.4.3, we studied another class of network dynamics, that in 
which there is `entanglement', in the sense that dynamics of 
some partial functions are influenced by other partial functions.
The structure of such entanglement, i.e., the mutual influence of different partial functions on
the rules for their dynamics, was analyzed.
In Secs.4.3.1 and 4.3.2, this entanglement was studied by
considering particular sets of initial conditions.  
It was shown that a perturbation
(representing an input to one partial function) circulates through a loop
in the network, and as a result, the network dynamics can switch to a new attractor.
The ability of this network to model a memory was discussed.

In conclusion, we have found that the FD system we have studied can form a hierarchically
entangled dynamical network.
We believe that such a system may be useful in modeling the spontaneous emergence of rules 
governing dynamics in biological system.

\subsection{Dynamical Network}

In general, a network consists of {\bf elements} and directed {\bf edges} between {\bf elements}.
{\bf Information} is transmitted through directed edges, 
and the internal state 
of an element (i.e., the value of the partial function)   
changes accordingly. 
The terminal of a given directed edge depends on the dynamics of this internal state.

In this paper, the function of a FD system is represented by a network 
according to the following prescription:

\begin{itemize}
\item An {\bf element} in the network consists of 
a connected subinterval on which the value of the function $f_n(x)$ is constant.
According to the results of our simulations, 
it appears that the formation of such constant subintervals is a general characteristic of 
FD systems, as discussed in Sec.3.

\item Directed {\bf edge}s are determined by the relations among the intervals induced by $f_n$. 
Specifically, if $f_n(A) \subseteq B$ is satisfied, there is a directed edge from
$B$ to $A$, represented as 
$B \rightarrow A$ in the graph of a network.

\item The {\bf internal state of the element} $A$ is defined as the value $f_n|_A(x)$.

\item The transmitted {\bf information} from $B$ to $A$ is the value $f_n|_B(x)$.
\end{itemize}

The shape of a network changes through the change of
its directed edges.
For example, the edge $B \rightarrow A$ changes to $C \rightarrow A$, 
if $g_n\circ f_n(A) \subseteq C$.  With this representation,
a dynamical network for a FD system can be formulated.\\

The advantage of a FD model lies in the simultaneous formation of elements 
and edges.  Through the FD,
elements and edges emerge from an infinite-dimensional 
functional space.
In most studies of dynamical networks, a fixed set of elements 
along with rules that govern the evolution of the edges are prescribed.\footnote{
This holds for all neural network models, and if also
holds for the models used in recent studies of networks \cite{small,barabasi,ito}.}
However, many biological (including cognitive and social) systems 
have the ability to generate rules governing their dynamics spontaneously.
As a simple example, consider signal transduction in a cell.
There, signal molecules
change the state of a cell, and play a role in determining the dynamics.
However, the results governing this function are not necessarily prescribed
by chemical reaction rules, but are determined by
the state of the cell,
which is realized through the reaction network dynamics.
It is true that investigation to elucidate the microscopic dynamics involving signal molecules
is important to understand such context-dependent functions of signal
molecules.  However,
it is also necessary to adopt an approach from a higher, 
macroscopic level, which can be represented by
a functional system.  Of course, the same can be said about
a cognitive systems consisting of neurons.

The FD approach is based on self-referential structure represented 
by $f_n\circ f_n$, which was introduced 
to understand how elements and rules are generated simultaneously
and interdependently.
The fixed point $f_n(x) = f_n\circ f_n(x)$ represents
a self-consistent condition in which operation by
the function  does not change the function itself.
This fixed point determines intervals on which the function is constant.
These intervals act as the basic elements in the network,
and drive other intervals.  
The constant intervals formed due to fixed points provide rules 
governing the dynamics of the other intervals.
The network thus generated can possess hierarchy and
entanglement.  

Hierarchy in a network is commonly observed in  biological
(as well as cognitive and social) systems.  In some cases, such 
hierarchy is simple, and the network possesses a tree structure.
In some other cases, the hierarchy is not simple, and 
the network is entangled; i.e., there exists
mutual dependence of the dynamics of elements.
In such a case, the network itself changes spontaneously.
Such network dynamics can also be changed with inputs.
Features of this kind that exist in real biological networks were also shown to exist
in our FD system in this paper. Considering the simplicity of our
model (1), the results found here give reason to believe 
that these features exist quite generally in network systems 
that maintain themselves through temporal evolution.
It is therefore important to
study how concepts formulated with regard to FD can be applied
to real biological network dynamics.

\subsection{Future Problems}

The limit function investigated here starts from a piecewise constant function $f_0$.
However, in a more general investigation, we also need to consider the case 
in which such initial functions are continuous.
With a continuous $f_0$, the limit function can consist of 
an infinite number of constant partial functions.
We believe that infinite hierarchy and infinite entanglement 
can be generated in this case. Characterization of the dynamics in a system with
such infinite hierarchy is a future problem.

The stability of a higher-order structure, such as
hierarchy and entanglement in a network, is an important problem in FD.
Though the theorem stated in Sec.3 is valid only for an $n$-independent partial function
(although we hope to generalize it in future),
we believe that attractors in a FD system are stable, 
because the slope of the generated map is $1-\e$.

By adding noise (i.e., an input to one or more partial functions)
with a finite magnitude, a transition between attractors can occur.
Consider a perturbation applied to the system by changing the value of $f_n|_{I^i}$.
As a result, any partial function $f_n|_{I^j}$ that satisfies 
$f_n|_{I^j} \subseteq I_i$ is influenced by this change in $f_n|_{I^i}$.
In this manner, applied noise is transmitted along edges in the network.
During this transmission, the noise 
1) will decrease due to the contracting nature of the generated map $g_n$,
but 2) it may cause a transition to a new attractor for each $f_n|_{I^i}$ when
the noise causes  $f_n|_{I^j}$ to switch to a new branch 
in the generated map. (Note that the generated map
is discontinuous.)
If a transition to a new attractors does not occur for any $n$, 
the influence of the noise eventually disappears, due to the contraction imparted by the 
generated map.\\

%
%
%
%
%

Considering the successive addition of inputs (noise), one can study 
how the inputs are `memorized' in the network.
Note, however, that there is a difficulty involved with the continuous addition of noise in our model.
Also, when noise is added to fixed points,
the network structure may be completely destroyed.\footnote{If the noise is applied instantaneously,
in most cases, the $n$-independent function around 
the fixed point changes to a new $n$--independent function around a new
fixed point, and the structure of the network is preserved.}  
Indeed, in a FD system with noise added continuously in time, 
if the magnitude of this noise is sufficiently large,the function 
becomes constant over all intervals.
One possible way to avoid this problem is to consider a
FD on a torus ($S^1$), i.e., a map on a circle.
In this case there exists a function that has no fixed points.
The study of FD on a circle is an interesting extension to be considered in the future.

\vspace{2cm}

\noindent{\bf Acknowledgment}

The authors acknowledge Y. Takahashi and T. Namiki for their discussion and mathematical suggestions.
They thank T. Ikegami and S. Sasa for stimulating discussions.
This work is partially supported by Grants-in-Aid for Scientific Research from
the Ministry of Education,
Science and Culture of Japan.


\begin{thebibliography}{999}

\bibitem{NKKI}
N. Kataoka, K. Kaneko ``Functional Dynamics I : Articulation Process'',
Physica D 138 (2000) 225-250.

\bibitem{NKKII}
N. Kataoka, K. Kaneko, ``Functional Dynamics II: Syntactic structure'',
Physica D 149 (2001) 174-196.

\bibitem{TKKN}
Y. Takahashi, N. Kataoka, K. Kaneko and T. Namiki, ``Function Dynamics'',
Japan J.Appl.Math. 18 (2001) 405-423.

\bibitem{YH}
M. Yamaguti and M. Hata
"Takagi function and its generalization"
Japan J. Appl. Math. 1 (1984) 186-199

\bibitem{Feigenbaum}
M.J. Feigenbaum,
"The universal metric properties of nonlinear transformations"
J. Stat. Phys. 21 (1979) 669

\bibitem{small}
D.J. Watts, ``Small Worlds", (Princeton Univ. Press, 1999)

\bibitem{barabasi}
R. Albert and A.L. Barabasi,
``Topology of evolving networks: Local events and universality",
Phys. Rev. Lett. 85 (2000) 5234

\bibitem{ito}
J. Ito and K. Kaneko,
` Spontaneous structure formation in a network of chaotic units with variable connection 
strengths", 
Phys. Rev. Lett., 88 (2002) 028701-1

\end{thebibliography}
\end{document}